\documentclass[a4paper,fleqn,usenatbib]{mnras}

\usepackage{newtxtext,newtxmath}
\usepackage[T1]{fontenc}
\usepackage{ae,aecompl}

\usepackage[dvipsnames]{xcolor}
\usepackage{graphicx}	
\usepackage{amsmath}	
\usepackage{BibDef}

\usepackage{ragged2e}

\usepackage{booktabs}

\usepackage{multirow}

\usepackage{mathrsfs}

\usepackage{comment}

\usepackage[normalem]{ulem}

\newcommand{\msun}{\ensuremath{\, \mathrm M{\sun{}}}}

\title[Partial disruption event rates]{Partial stellar tidal disruption events and their rates}

\author[E. Bortolas  et al.]{
Elisa Bortolas$^{1,2}$\thanks{E-mail: elisa.bortolas@unimib.it}, Taeho Ryu$^{3,4}$, Luca Broggi$^{1,2}$ and Alberto Sesana$^{1,2}$
\\
$^{1}$ Dipartimento di Fisica ``G. Occhialini'', Università degli Studi di Milano-Bicocca, Piazza della Scienza 3, I-20126 Milano, Italy\\
$^2$ INFN, Sezione di Milano-Bicocca, Piazza della Scienza 3, I-20126 Milano, Italy\\
$^3$ Max Planck Institute for Astrophysics, Karl-Schwarzschild-Strasse 1, 85748 Garching, Germany\\
$^{4}$ Physics and Astronomy Department, Johns Hopkins University, Baltimore, MD 21218, USA\\
}

\date{Accepted XXX; Received YYY; in original form ZZZ}

\pubyear{2023}

\begin{document}
\label{firstpage}
\pagerange{\pageref{firstpage}--\pageref{lastpage}}
\maketitle

\begin{abstract}
Tidal disruption events (TDEs) of stars operated by massive black holes (MBHs) will be detected in thousands by upcoming facilities such as the \textit{Vera Rubin Observatory}. In this work, we assess the rates of standard \textit{total} TDEs, destroying the entire star,  and \textit{partial} TDEs, in which a stellar remnant survives the interaction,  by solving 1-D Fokker-Planck equations. Our rate estimates are based on a novel definition of the loss cone  whose size is commensurate to the largest radius at which partial disruptions can occur,  as motivated by relativistic hydrodynamical simulations. 
Our novel approach unveils two important results. First, partial TDEs can be more abundant than total disruptions by a factor of a few to a few tens. 
Second, the   rates of complete stellar disruptions can be overestimated by a factor of a few to a few tens if one neglects partial TDEs, as we find that many of the events  classified as total disruptions in the standard framework are in fact partial TDEs. 
Accounting for partial TDEs is particularly relevant for galaxies harbouring a nuclear stellar cluster featuring many events coming from the empty loss cone.
Based on these findings, we stress that  partial disruptions should be considered when constraining the luminosity function of TDE flares; accounting for this may reconcile the theoretically estimated TDE rates with the observed ones.
\end{abstract}

\begin{keywords}
transients: tidal disruption events -- galaxies: kinematics and dynamics -- stars: kinematics and dynamics -- methods: numerical --  black hole physics
\end{keywords}



\section{Introduction}\label{sec:intro}
Almost all massive galaxies harbour a massive black hole (MBH, with mass $\gtrsim 10^5\msun{}$) at their centre  \citep{Kormendy2013}. Occasionally, weak gravitational encounters between stellar objects can place some stars on nearly parabolic orbits around the MBH: if their pericentre distance gets smaller than the so-called ``\textit{tidal disruption radius}'', the tidal forces generated by the MBH  completely dominate the self-gravity across the entire star, resulting in a total tidal disruption event (TTDE, \citealt{Hills1988}). When this happens, roughly half the mass of the stellar debris remain bound to the MBH; as the bound debris return and ultimately fall onto the MBH, they produce an extremely bright flare \citep{Rees1988, Lodato2009}.

If the pericentre distance of the passing star is small but  greater than the tidal disruption radius, \textit{partial tidal disruption events} (PTDEs) can occur, in which a stellar remnant survives the disruption after losing  a fraction of its mass. Like TTDEs, also in case of PTDEs some fraction of the debris would return to the MBH. But the mass fallback rate is lower at peak and declines more rapidly than that for TTDEs of the same star \citep[e.g.,][]{Guillochon+2013,Goicovic+2019,Ryu+2020c}. PTDEs  affect the thermodynamic and hydrodynamic state and the orbital energy of the stellar remnant; upon disruption, the remnant is spun-up and gets  hotter than an ordinary  star of the same mass \citep{Ryu+2020c}. Furthermore, the  orbital energy can significantly vary \citep[e.g.,][]{Manukian+2013,Ryu+2020c} due to asymmetric mass loss and tidal excitation, possibly resulting in a momentum kick. If partially disrupted stars remain gravitationally bound to the MBH, they can in principle undergo repeated PTDEs \citep[][]{Ryu+2020c}.

To date, roughly 100 TDE candidates have been detected \citep{Gezari2021}, and their observations are already proving to be valuable tools to infer the properties of the central MBH and the destroyed star \citep[e.g.][]{Mockler2019,Ryu+2020e,Mockler2021}. The number of observed TDEs  is bound to dramatically increase thanks to upcoming detections by ongoing \citep[e.g., eROSITA][]{Sazonov+2021} and future facilities (as the \textit{Vera Rubin Observatory}, \citealt{BricmanGomboc2020},  and ULTRASAT, \citealt{Sagiv2014}) allowing us to test their  predicted event rates. 

Most of the  theoretical estimates of the TDE rates rely on the so-called loss cone theory. With this approach, the evolution of stellar orbits, mainly driven by two-body encounters, is described as a diffusion process in phase space. In this formalism, one can estimate the rate at which stars enter a small region of parameter space whithin which they can undergo a TDE (the ``loss-cone'') by solving a static or time-dependent Fokker-Planck equation. In general, assuming isotropic and spherically symmetric stellar systems, theoretical models predict that the event rates are higher in smaller galaxies, and the overall  per-galaxy event rate is $\sim 10^{-4}-10^{-5}~{\rm yr}^{-1}$ (e.g. \citealt{SyerUlmer1999,Magorrian1999,Wang2004,Stone2016,Bortolas2022}; see also the recent work by \citealt{Teboul2022}).
Few of the observed TDE candidates present features that may be more consistent with a PTDE nature. Some of them were detected with relatively faint light curves which decayed rapidly (e.g., AT2019qiz, \citealt{Nicholl+2020}, and iPTD16fnl,  \citealt{Blagorodnova+2017}), as expected from the mass fallback rate for PTDEs that peaks at a lower values and decays faster than that for TTDEs \citep{Ryu+2020c}. Some nuclear transients revealed periodic bursts (e.g., ASASSN--14ko, \citealt{Payne+2021,Liu2023}, eRASSt J045650.3--203750, \citealt{Liu2023b}, AT 2018fyk, \citealt{Wevers2023}, RX J133157.6--324319.7, \citealt{Malyali2023}, and quasi periodic eruptions, \citealt{Miniutti+2019,Giustini+2020,Arcodia+2021,Franchini2023}), which may be caused by repeated partial disruption events. In particular, the population of repeating events has grown recently. Although the nature of most of the  known repeating events remains elusive, the possible association of them with repeated PTDEs adds to the importance of an accurate rate estimate for PTDEs.

TDE rate estimates have been refined throughout the last years thanks to advances in their dynamical modelling. However, most of the previous theoretical work  has focused exclusively on TTDEs. Nonetheless, PTDEs have a larger encounter cross-section, implying that they can be more frequent than TTDEs \citep{Krolik+2020}; since PTDEs involving the same star can occur multiple times, their overall rate can be further enhanced. Even if PTDEs are likely intrinsically dimmer than TTDEs, they can still be detectable if the subject star undergoes a significant mass loss, while the unprecedented capabilities of future surveys as LSST can in principle disclose PTDEs even if the subject star undergoes little mass loss \citep{BricmanGomboc2020}. Furthermore, the mass accreted during PTDEs can substantially impact the growth of the central MBH. It is therefore important to shed light on the rates of PTDEs and the typical galaxies that may host them. So far, there have been only a few  attempts to do so, and in those, somewhat specific astrophysical situations were considered. \citet{ChenShen2021} attempted to analytically estimate the volumetric PTDE rate assuming stars are in (quasi-)steady states. They found that the volumetric PTDE rate for MBHs with mass $\lesssim 10^{6}M_{\odot}$ is comparable to that of TTDEs, but becomes dominant for larger masses. More recently, using direct $N-$body simulations of a cluster harbouring an MBH of $10^{6}M_{\odot}$, \citet{Zhong+2022} compared rates of TTDEs with those of PTDEs; they found that the number of PTDEs is roughly 75\% higher than the simple analytic estimate based on the size of the encounter cross-section, owing to the fact that single stars can undergo repeated PTDEs. 

In this paper we develop a novel modelling strategy to systematically investigate the rate of TTDEs and PTDEs; our framework is based on the Fokker-Planck approach. We address a  wide range of MBH masses ($10^{5}M_{\odot}\leq M_{\bullet}\leq 10^{7.5}M_{\odot}$) embedded in a range of stellar host. Sec.~\ref{sec:methods} provides a detailed description of our methodology; in Sec.~\ref{sec:MW} we specialise our treatment to a Milky-Way-like galaxy nucleus, while in Sec.~\ref{sec:multimodel} we explore PTDE and TTDE rates in a broader set of models. Finally,  Sec.~\ref{sec:discussion} summarises our findings and discusses   their astrophysical implications.


\section{Modelling of the partial disruption rates}\label{sec:methods}

\subsection{Total and partial disruption radius}

\begin{figure}
\includegraphics[ width=0.45\textwidth]{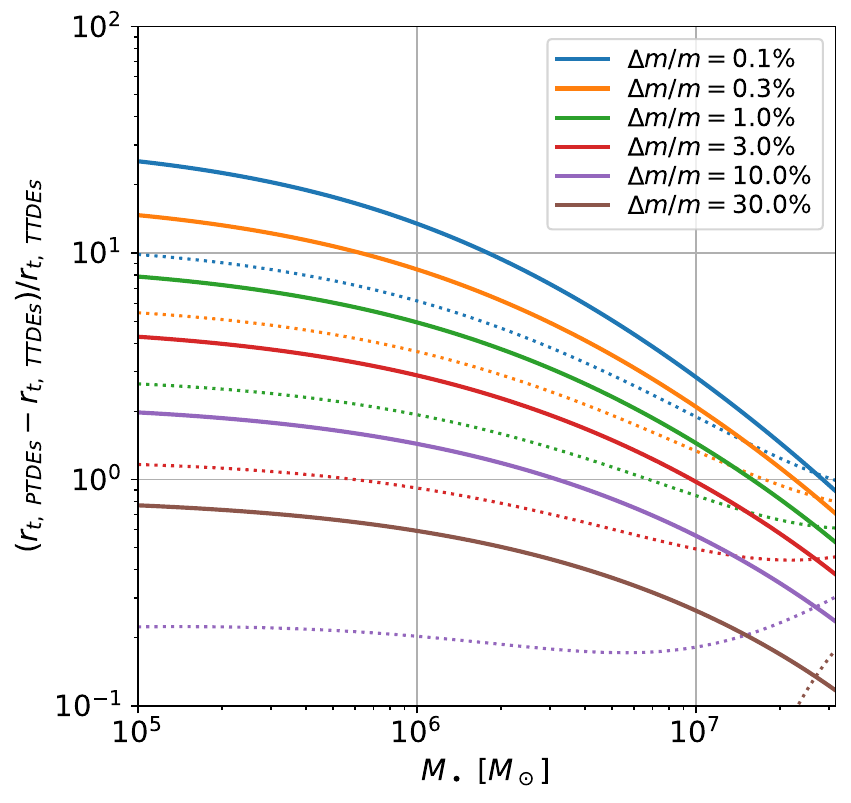}
    \caption{Relative enhancement of the tidal disruption radius $r_t$  for PTDEs compared with the same value associated with TTDEs as a function of the MBH mass, for different values of the mass stripped during a PTDE, $\Delta m/m$. All lines assume the disrupted star weighs 1~$\msun$. The solid lines consider the TTDE radius to be computed adopting $\eta$ from Eq.~\ref{eq:eta_ftde} (so that the plotted quantity is in fact $(\Delta m/m)^{-1/\zeta}-1$); the dotted lines assume instead $\eta=1$ for TTDEs, so that the plotted quantity is equal to $\eta_{\rm PTDE}-1$.}
    \label{fig:radius_enhancement}
\end{figure}

In this paper, we define as \textit{PTDE} an event that does not completely disrupt a star, but only strips a fraction of its mass leaving behind a stellar core; we also include in the PTDE count the final disruptions of stars that already underwent at least one PTDE (differently from \citealt{Zhong+2022}). We name \textit{TTDEs}  the events associated with stars that never underwent a PTDE before and get completely destroyed in a single event. In order to quantitatively define the two, we need to specify their associated tidal disruption radii. It is custom to express the tidal disruption radius as
\begin{equation}\label{eq:rt}
r_t = \eta \left(\frac{M_\bullet}{m}\right)^{1/3} R_\star
\end{equation}
where $m$ and $R_\star$  are the stellar mass and radius, $M_\bullet$ is the MBH mass and $\eta\sim1$ is a form factor which is generally varied depending on the internal structure of the star  and the impact of relativistic effects. \citet{Ryu2020eta} provide an empirical value for $\eta$ based on numerical simulations, that defines the radius below which a star is totally disrupted:
\begin{equation}\label{eq:eta_ftde}
\begin{split}
    \eta_{\rm TTDE}(m, M_\bullet) = \left(0.80+0.26\sqrt{\frac{M_\bullet}{10^6\msun}}\right)\times \\
    \left(\frac{1.47 + \exp{\left(\frac{m/\msun -0.669}{0.137}\right)}}{1+2.34\exp{\left(\frac{m/\msun -0.669}{0.137}\right)}}\right).
\end{split}
\end{equation} 
This prescription matches relatively well the results obtained in different works  \citep{Law-Smith2019, Law-Smith2020}.

The PTDE radius, instead, is a larger radius that does not ensure the complete disruption of the star.
In case of PTDEs of a $1\msun$ middle-age main-sequence star, \citet{Ryu+2020c,Ryu+2020d} use numerical simulations to show that the value of $\eta_{\rm PTDE}$ depends on the relative fraction of the stellar mass that is stripped in the event, $\Delta m/m$. In particular\footnote{All $\log$ symbols in the following equations refer to logarithms with base 10.} 
\begin{equation}\label{eq:eta_ptde}
\begin{split}
    \eta_{\rm PTDE}(\Delta m/m, M_\bullet) = \eta_{\rm TTDE}(m, M_\bullet) \times (\Delta m/m)^{-1/\zeta}, \\
\end{split}
\end{equation} 
where $\log\zeta= 3.15\times10^{-8}[\log(M_\bullet/\msun)]^{8.42} + 0.3$. 

Note that in many applications, the TTDE radius is computed by simply setting $\eta=1$; for comparison, in our computation we will also explore how the event rates vary if we simply set $\eta=1$ for TTDEs. The relative enhancement in the PTDE radius with respect to the TTDE radius is shown in Fig.~\ref{fig:radius_enhancement}.

\subsection{Loss cone theory}\label{sec:loss_cone_theory}
The means to estimate the rate of stellar disruptions in a given astrophysical system rely on the loss-cone theory. Here, we do not go into the fine details of the theory; instead, we focus on the concepts that are crucial for our implementation, and we refer the reader to the comprehensive review by \citet[][and references therein]{Stone2020}
for more details. For now, we present the general framework without specifying whether we are focusing on TTDEs or PTDEs, but assuming an arbitrary value for $r_t$. We specialize our treatment to PTDEs versus TTDEs in Sec.~\ref{sec:ptde_theory}.

We will work under the assumption of spatial isotropy and spherical symmetry of the host galaxy. Let's consider a system that is described by a density profile $\rho(r)$ that, together with the presence of a central MBH of mass $M_\bullet$, defines the stellar distribution function $f(E)$ that only depends on the specific  energy $E$ of a given star. Furthermore, it is custom to define the variable 
\begin{equation}\label{eq:R}
\mathcal{R}(E)=j^2/j^2_{\rm circ}(E),
\end{equation}
where $j$ is the specific  angular momentum of a star and $j_{\rm circ}(E)$ is the specific angular momentum of a star on a circular orbit with energy $E$; note that in the Keplerian limit, $\mathcal{R}\approx 1-e^2$, with $e$ eccentricity of a stellar orbit. Obviously, a necessary condition for a star to be destroyed by the MBH is to have a pericentre smaller than $r_t$, or equivalently an angular momentum smaller than
\begin{equation}
j_{\rm lc}=\sqrt{2GM_\bullet r_t}, 
\end{equation} 
which corresponds to a value of $\mathcal{R}$  equal to
\begin{equation}\label{eq:Rlc}
\mathcal{R}_{\rm lc}(E)=j^2_{\rm lc}/j^2_{\rm circ}(E).
\end{equation}
The loss cone is defined as the region of phase space with $j<j_{\rm lc}$. Stars can get in and out from the loss cone due to the effect of stellar two-body relaxation, that ultimately sets the event rate. 

Near the loss cone, relaxation in angular momentum is more efficient than relaxation in energy; this is due to the fact that small perturbations are sufficient to perturb $j$ of stars on very low angular momentum orbits by an amount of the order of $j$ itself. With this notion, it is possible to study the system via the one-dimensional Fokker-Planck equation that only considers diffusion in angular momentum and neglects energy diffusion. First of all we need to define
\begin{equation}
    N(E,\mathcal{R},t) = 4 \pi T_{\rm orb}(E)j^2_{\rm circ}(E) f(E, \mathcal{R}),
\end{equation}
representing the number of stars per unit interval of $E$ and $\mathcal{R}$; since the orbital period is weakly dependent of $\mathcal{R}$, it is reasonable to average its value over all available $\mathcal{R}$ and consider it as a function of $E$ only. The 1-D Fokker-Plank equations thus read
\begin{equation}\label{eq:fokker-planck}
    \frac{\partial N(E, \mathcal{R},t)}{\partial t} = \frac{\partial \mathcal{F}(E,\mathcal{R}, t)}{\partial \mathcal{R}}; \quad \mathcal{F}(E,\mathcal{R}, t) \approx \mathcal{D}(E) \mathcal{R} \frac{\partial N(E, \mathcal{R},t)}{\partial 
 \mathcal{R}},
\end{equation}
where $\mathcal{F}$ is the flux of stars towards the loss cone, while $\mathcal{D}(E)$ is the angular momentum diffusion coefficient that sets the relaxation rate and depends on the moments of the distribution function \citep[see e.g. ][]{Merritt2013}; note that $\mathcal{D}^{-1}(E)$ yields a good approximation for the relaxation timescale. The boundary conditions for these equations read
\begin{equation}
    \frac{\partial N(E, \mathcal{R},t)}{\partial \mathcal{R}}\bigg\rvert_{\mathcal{R}=1}=0; \  N(E, \mathcal{R}_{\rm lc},t)-\alpha(E) \mathcal{R}_{\rm lc}\frac{\partial N(E, \mathcal{R},t)}{\partial \mathcal{R}}\bigg\rvert_{\mathcal{R}=\mathcal{R}_{\rm lc}}=0
\end{equation}
where 
\begin{equation}\label{eq:alpha}
\alpha=\sqrt{q^2+q^4}    
\end{equation}
and $q$ is the so-called loss-cone filling factor
\begin{equation}\label{eq:q}
    q(E) = \frac{\mathcal{D}(E)T_{\rm orb}(E)}{\mathcal{R_{\rm lc}}(E)}.
\end{equation}
A well-known and widely used solution for Eq.~\eqref{eq:fokker-planck} is the quasi-steady-state solution derived by \citet{Cohn1978}, that corresponds to a relaxed distribution in angular momentum in a system featuring a loss-cone associated to stellar disruptions. \citet{Cohn1978} derive a solution for $f$ (or equivalently, $N$) close to the loss cone: for each given energy, 
\begin{equation}\label{eq:cohnkulsrud}
 f(E, \mathcal{R}, t) \approx \begin{cases}
			A \ln(\mathcal{R}/\mathcal{R}_0) & \text{if $\mathcal{R_{\rm0}<R<R_{\rm lc}}$}\\
            0 & \text{if $\mathcal{R<R_{\rm 0}}$},
		 \end{cases}
\end{equation}
where the normalisation constant $A$ can be expressed in terms of the average value of $f$ and it is derived below, while
\begin{equation}\label{eq:R_0}
    \mathcal{R}_0 = \mathcal{R}_{\rm lc} e^{-\alpha}
\end{equation}
is the zero of the aforementioned equation.
Following the existing literature, we assume this solution holds inside the loss-cone  even if the distribution of stars at pericenter is set by the solution of the local, non orbit-averaged Fokker-Planck equation that holds there (Broggi et al., in preparation). Therefore, in this work $\mathcal{R}_0$ is the minimum value of the angular momentum inside the loss-cone below which no star exists.\footnote{Although our methodology for computing the distribution of stars in the loss cone as a function of $\mathcal{R}$ is standard and common to many recent papers \citep[e.g.][]{MerrittWang2005,Teboul2022}, a more accurate approach would require to adopt the non-orbit averaged solution for the distribution function inside the loss cone, as presented in \citet{Strubbe2011} and \citet[][their eq. 6.54]{Merritt2013}. We are currently assessing the impact of adopting the non-orbit averaged solution on the PTDE rates, and we anticipate that the number of TTDEs we find in the present paper may be slightly underestimated when most of the events come from $q\approx 1$. This result is about to be presented in Broggi et al., in prep.}

\subsubsection{Full and empty loss cone regimes}\label{sec:emptyfulllc}
Before proceeding further, it is worth giving a more physical intuition of the quantities described so far and to introduce the concept of empty and full loss cone. First of all, it is important to highlight that
\textit{the loss-cone filling factor $q(E)$ (Eq.~\ref{eq:q}) can be interpreted as the ratio between the orbital period $T_{\rm orb}(E)$ and the typical timescale over which a star on the edge of the loss-cone is deflected out of it by relaxation processes. } $\mathcal{D}^{-1}(E)$ can indeed  be interpreted as the timescale over which an object on a circular orbit changes its squared angular momentum by an amount of the order of itself; if the object is instead on a much smaller angular momentum (more eccentric) orbit, the timescale over which its squared angular momentum changes by the order of itself is reduced by an amount equal to $\mathcal{R}=j^2/j^2_{\rm circ}$. This implies $\mathcal{R}_{\rm lc}(E)\mathcal{D}^{-1}(E)$  represents the typical time an object on an (eccentric) orbit at the edge of the loss cone requires to get out of it due to two-body stellar interactions. With these notions, we can divide the space about the MBH into two regions. 

Stars far away from the MBH are characterised by $q\gg1$, and they are said to be in the  \textit{full loss-cone regime}. This simply means that stars on the edge of the loss-cone can enter and exit it many (on average $\approx q$) times over a single orbital period, implying that even if a star has an instantaneous pericentre equal to (or smaller than) $r_t$, it is not guaranteed to be disrupted as it is likely to be scattered into a ``safer orbit'' via relaxation processes before reaching the pericenter. In this region, the loss-cone is said to be \textit{full} as relaxation is very efficient in continually bringing new stars into it, and every star lost in a TDE is statistically replaced by a new one.

Closer to the MBH $q\ll1$ and stars inhabiting this region are said to be in the \textit{empty loss cone regime}. This  means that stars on loss cone orbits typically need many ($\approx q^{-1}$) orbital periods to get out of it as a result of relaxation. This also implies that once a star enters the loss-cone, it is very likely to undergo a disruption at its first pericentre. Destroyed stars coming from the empty loss cone are not statistically immediately replaced by others  -- thus the loss cone is said to be \textit{empty}.

With this distinction in mind, it is easier to interpret the meaning of Eq.~(\ref{eq:cohnkulsrud}, \ref{eq:R_0}). At any given time, stars \textit{near the loss cone} undergo a Brownian wandering for their value of $\mathcal{R}$ (while their $E$ remains constant to a good approximation). The width of these Brownian steps in $\mathcal{R}$ crucially depends on   $q(E)$: after one entire orbit, the value of $\mathcal{R}$ of a star on the edge of the loss-cone would change by an amount $\sim q\mathcal{R}$.  
Since for stars in the empty loss cone regime $q\ll1$, the variation of angular momentum per orbit is tiny and they generally get disrupted when jumping from just outside the loss cone to an orbit with $\mathcal{R}=(1-q)\mathcal{R}_{\rm lc}\approx \mathcal{R}_{\rm lc}$. Accordingly,  $\mathcal{R}_0\approx \mathcal{R}_{\rm lc}$ (Eq.~\ref{eq:R_0}) so that $f(\mathcal{R})$ in the loss-cone is close to a step function at $\mathcal{R}_{\rm lc}$. In the full loss cone regime, instead, stars experience along each orbit jumps in angular momentum that are statistically 
larger than the instantaneous value of their $\mathcal{R}$. This implies that stars on the loss cone boundary can be destroyed at all available values of $\mathcal{R}<\mathcal{R}_{\rm lc}$ since $\mathcal{R}_0\approx0$, and stars have a nearly flat distribution in $\mathcal{R}$ (the limit of Eq.~\ref{eq:cohnkulsrud} if $\mathcal{R}_0\rightarrow0$). In this situation, stars can have all possible values of pericentre passages when they get destroyed -- some can be accreted inside the MBH horizon without yielding an observable signal.
Obviously, in reality a system is characterised by all possible values of $q(E)$, and which regime between the two is the dominant one depends on the value of $E$ (and the associated $q$) at which the TDE flux is maximum.

\subsubsection{The disruption rates}

In the 1-D approximation we are using, one can use as a reference the $\mathcal{R}$-averaged distribution function $\bar{f}(E)$ and associate to its value the flux corresponding to the \citeauthor{Cohn1978} function with the same average value, namely
\begin{equation}
	\bar{f} = A \int_{\mathcal{R}_{\rm lc}}^1 d\mathcal{R} \ln \frac {\mathcal{R}}{\mathcal{R}_0} = A\left[ (\alpha - 1)(1-\mathcal{R}_{\rm lc}) - \ln(\mathcal{R}_{\rm lc}) \right]
\end{equation}
so that the flux inside the loss-cone contributed by an energy bin $dE$ is given by
\begin{equation}\label{eq:TDEflux}
    \mathcal{F}(E) dE =  \frac{\bar{f}(E)\mathcal{D}(E)}{(\alpha - 1)(1-\mathcal{R}_{\rm lc}) - \ln(\mathcal{R}_{\rm lc})} dE.
\end{equation}
Finally, the total flux  of disruptions is simply the integral of the above flux for all available energy values
\begin{equation}\label{eq:total_flux}
   \Gamma  = \int\mathcal{F}(E) dE.
\end{equation}

\subsection{Non-repeating PTDE rates}\label{sec:ptde_theory}

Generally, the loss-cone theory and thus the \citeauthor{Cohn1978} solution is implemented only accounting for TTDEs, and adopting the corresponding value for $r_t$. Here we argue that this solution should rather be applied adopting the PTDE radius. 

After a star undergoes a  PTDE, its subsequent evolution can proceed in different ways: if the star comes from the empty loss-cone region, it can be entirely consumed through a series of repeated PTDEs,  
or the (possibly repeated) PTDE can significantly deflect the stellar orbit so that the star is scattered on a completely different trajectory before being disrupted. If  the star has $q\sim1$, its orbit may change appreciably after a PTDE, so that the star either gets destroyed on the next passage(s) or gets pushed away from the loss cone \citep[e.g.][]{Cufari2023}.  If the star comes from the full loss-cone, its orbit following the PTDE would be substantially deflected by relaxation processes so that it will become statistically indistinguishable from other stars in the full loss cone with the same energy. Either way, the theory developed so far for modelling TTDEs (and described in Sec.~\ref{sec:loss_cone_theory}) is suited for computing TTDE \textit{plus} PTDE rates instead, as the outcome of the process is virtually the same as in the traditional TTDE-only treatment (i.e. the star is lost from the system, or it ends up far away from the  the loss cone; even if this process might not be as fast as in the traditional treatment that considers only TTDEs, we can still consider it to be instantaneous as the star is lost from the loss cone within a timescale much smaller than the relaxation timescale of the host system). Below we also argue that the same traditional treatment yields an overestimated TTDE rate, if it is applied assuming that only TTDEs can occur. For now, we do not account for the complication that each star can undergo a series of subsequent PTDEs when computing the event rates; we focus on this aspect in the next section.

\begin{figure}
\includegraphics[ width=0.45\textwidth]{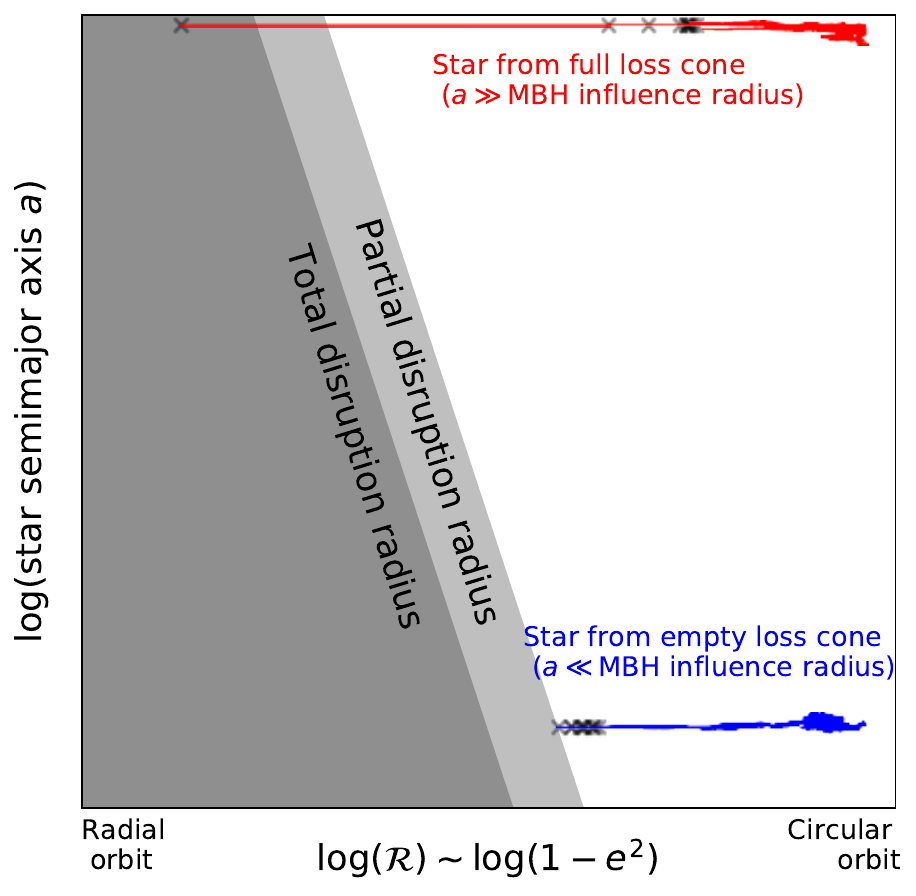}
    \caption{This cartoon qualitatively illustrates the behaviour of tidally disrupted stars coming from the full (red) or the empty (blue) loss cone. In particular, the image depicts the Brownian wandering of the semimajor axis $a$ and normalised angular momentum $\mathcal{R}$  of two different stars respectively coming from the full and empty loss cone; the light and dark grey areas highlight the regions in which respectively PTDEs and TTDEs can occur, with their right-most boundary corresponding to their associated disruption radius. Along the two colored stellar tracks, the values of $a$ and $\mathcal{R}$ associated with the last 10 pericenters before the (total or partial) disruption are marked with a black $\times$. The star coming from the full loss cone undergoes huge jumps in its value of $\mathcal{R}$ (much larger than the value of $\mathcal{R}_{\rm lc}$ at that semimajor axis) within each orbital period, and ends up being disrupted with a radius much smaller than the first available radius at which PTDEs can occur. On the other hand, the star coming from the empty loss cone undergoes very tiny jumps in $\mathcal{R}$ during each orbital period (much smaller than the value of $\mathcal{R}_{\rm lc}$ at that semimajor axis) so that it nearly adiabatically approaches the first available radius at which a PTDE can occur and gets destroyed there.}
    \label{fig:cartoon}
\end{figure}

Let us assume that PTDEs can occur below a given specific value of $r_t$; with this in mind, we can use Eq.~\ref{eq:cohnkulsrud} to know what is the probability of stars to be disrupted at each $\mathcal{R}$ inside this newly defined PTDE loss cone. In the full loss cone limit, all values of $\mathcal{R}$ in the loss-cone are equally probable, so that the ratio of TTDEs to PTDEs can be simply estimated through the ratio between their respective radii. In the empty loss-cone limit, instead, all stars tend to be destroyed at $\mathcal{R}_{\rm lc}$, implying that all stars would undergo a PTDE, while TTDEs are completely suppressed; this simply comes from the fact that stars approach the loss cone boundary with very small steps in $\mathcal{R}$, and they can only be (partially) destroyed at the first available radius that allows this to occur, while it is virtually impossible for stars to reach the smaller values of $\mathcal{R}$ at which TTDEs can occur (the concept is illustrated in Fig.~\ref{fig:cartoon}). This implies that systems in which TDEs mainly come from the empty loss cone would have nearly all stars undergoing PTDEs, and their TTDE rates can be very small in comparison.

More quantitatively, for a fixed value of $E$, we can use Eq.~\ref{eq:cohnkulsrud} to compute the expected ratio of TTDEs to PTDEs. If $r_t$  is the largest PTDE radius that guarantees an event to occur (with its associated PTDE $\mathcal{R}_{\rm lc}$), and $r_t/\kappa$  is the radius below which only TTDEs occur ($\kappa>1$), the fraction of TTDEs to   the global event rate (PTDEs \textit{plus} TTDEs) can be estimated as\footnote{In the two equations below, the dependence of all quantities on energy is not explicitly written. }
\begin{equation}
\delta_{\rm TTDE} = \frac{\int_{\mathcal{R}_0}^{\mathcal{R}_{\rm lc}/\kappa}d\mathcal{R} \ln(\mathcal{R}/\mathcal{R}_0) \, \theta(\mathcal{R}_{\rm lc}/\kappa-\mathcal{R}_0)}{\int_{\mathcal{R}_0}^{\mathcal{R}_{\rm lc}}d\mathcal{R} \ln(\mathcal{R}/\mathcal{R}_0)}
\end{equation}
where $\theta$ is the step function, that is one if its argument is positive and zero otherwise. We stress that all the relevant quantities ($\alpha$, $\mathcal{R}_0$ etc.) in this framework should be evaluated adopting the PTDE value of $r_t$.
The above equation can be evaluated to be 
\begin{equation}\label{eq:delta_ttde}
\delta_{\rm TTDE} = \frac{1}{\kappa}\left[1 + \frac{(\kappa-1)\,e^{-\alpha} - \ln \kappa}{\alpha + e^{-\alpha} - 1} \right] \,  \theta(\alpha-\ln \kappa)
\end{equation}
that goes to zero with continuity as $\alpha \to \ln \kappa$. As expected, $\delta_{\rm TTDE} \rightarrow 1/\kappa$ if $q\gg1$ ($\alpha \gg1$), i.e. in the full-loss cone limit, while it becomes 0  through the $\theta$ function in the empty loss cone regime. Finally, in the framework described so far, Eq.~\eqref{eq:total_flux} gives the total rate of  PTDEs plus TTDEs; the rate associated to TTDEs only can be computed as 
\begin{equation}\label{eq:flux_TTDE}
   \Gamma_{\rm TTDE}  = \int \delta_{\rm TTDE}(E) \mathcal{F}(E) dE, 
\end{equation}
while the rate of PTDEs is
\begin{equation}\label{eq:flux_PTDE}
   \Gamma_{\rm PTDE}  = \int  \delta_{\rm PTDE}(E) \mathcal{F}(E) dE,
\end{equation}
where $\delta_{\rm PTDE}(E) = 1-\delta_{\rm TTDE}(E)$.

\subsection{Accounting for repeating PTDEs}\label{sec:ptde_rate}

The strategy detailed in the previous section allows to estimate the rate of PTDEs assuming that only one PTDE per star can happen; however,  a single star can undergo multiple PTDEs before being scattered out of the loss cone or being completely disrupted. 

Stars coming from the \textit{full loss cone} would statistically undergo only a single PTDE; this is because they are very likely to be deflected out of the loss cone along the orbital period that follows their first PTDE. 
Stars in the \textit{empty loss cone} regime, instead, are more likely to undergo a series of PTDEs. As detailed in the previous section,  they tend to remain on the same orbit for $q^{-1}$ orbital periods before getting scattered out, so that in principle they can undergo $q^{-1}$ PTDEs. This latter estimate relies on the assumption that \textit{ the stellar orbit is not significantly altered during a PTDE, i.e. there is no significant velocity kick associated with the PTDE}. In fact, the orbit may be perturbed at the pericentre, but the magnitude of the velocity kick associated with a PTDE remains very debated. Following  \citet{Manukian+2013} and \citet{Zhong+2022}, the velocity kick at pericentre for solar-type stars would be $v_k = (0.0745 + 0.0571\beta^{4.539})\times 617.7$ km s$^{-1}$; here  $\beta = (M_\bullet/m)^{1/3}R_\star/r_p$ where $r_p$ is the stellar pericentre.\footnote{This equation is formally valid only for $\beta>1$, but we extrapolate this trend for lower values.} By setting $\beta = 1/\eta_{\rm PTDE}$, we find that $v_k$ is relatively small ($\approx 50$ km s$^{-1}$) unless $\Delta m/m>0.1$. As we will see below, repeated PTDEs are more likely to occur for small $\Delta m/m$, implying that the PTDE velocity kick may have a small overall effect on our computed rates; thus, we neglect it in our following analysis, and we defer the detailed treatment of such aspect to a forthcoming paper.

A further relevant point for our PTDE computation  is that the number of PTDEs that can be experienced by a single star is necessarily limited by its mass: if the star loses a mass $\Delta{m}$ at each passage and $m$ represents the  mass of the star prior to the first  PTDE, no stellar mass will be left after a number of PTDEs that is nearly $n_p\approx m/\Delta m$.
Since repeated PTDEs reduce the total mass of the stellar remnant, the stellar radius would also be affected. Yet, the  change in the stellar radius associated with PTDEs is not easy to compute as it would depend on the time the star needs to settle back on the main sequence compared with its orbital period. Here we make the simplifying assumption that the value of $r_t$ associated with a partial disruption does not vary even if a star undergoes a series of PTDEs. In any case, our overall results are always presented both accounting or not for the possibility of repeated PTDEs -- so that the latter can be considered to be an upper limit for the PTDE rate.

All the  above considerations bring us to estimate the number of  \textit{repeating} PTDEs that can be undergone by the same star as 
\begin{equation}
  N_{\rm rep} (\mathcal{R},E) = \begin{cases}
			\min \left[\frac{1}{q(E)}, \left(\frac{\Delta m}{m} (r_p(\mathcal{R},E))\right)^{-1}\right] & \text{if $\mathcal{R_{\rm lc}/\kappa<R<R_{\rm lc}}$},\\
            0 & \text{otherwise},
		 \end{cases}
\end{equation}
where $\tfrac{\Delta m}{m} (r_p(\mathcal{R},E))$ is a function that, given a periapsis 
 $r_p$ of a star (obviously determined through its values of $\mathcal{R},E$), converts it into a value of $\Delta m / m$ through the algebraic inversion of Eq.~\eqref{eq:rt} with $r_p$ used instead of $r_t$ and $\eta$  defined in Eq.~\eqref{eq:eta_ptde}. Obviously, $N_{\rm rep}=0$ if $r_p$ is smaller than the radius at which TTDEs can occur or if it is larger than the maximum radius at which PTDEs are allowed to happen.

The overall number of repeating PTDEs can be  estimated as 
\begin{equation}\label{eq:flux_RPTDE}
   \Gamma_{\rm rep }  = \int dE \mathcal{F}(E) \delta_{\rm rep }(E)
\end{equation}
where (dropping for simplicity the dependence of all shown variables on $E$)
\begin{equation}\label{eq:delta_rep}
    \delta_{\rm rep }=\frac{\int^{\mathcal{R}_{\rm lc}}_{\mathcal{R}_{\rm lc}/\kappa}d\mathcal{R} \ln(\mathcal{R}/\mathcal{R}_0)\, N_{\rm rep }(\mathcal{R}) \, \theta(\mathcal{R}_{\rm lc}/\kappa-\mathcal{R}_0)}{\int_{\mathcal{R}_0}^{\mathcal{R}_{\rm lc}}d\mathcal{R} \ln(\mathcal{R}/\mathcal{R}_0)}.
\end{equation}

To avoid confusion, in the rest of the paper we will always refer to the \textit{non repeating} rates as our fiducial rates, and explicitly state when we refer to the repeating PTDE rates.

\subsection{Methods and numerical approach}

In order to specialize our treatment to a given astrophysical system, we set up a model (as detailed below) and we compute all the relevant quantities for performing our rate estimates relying on the \textit{Phaseflow} Fokker-Planck integrator \citep{Vasiliev2017},  part of the AGAMA toolkit \citep[][]{Vasiliev2019}. This very efficient integrator handles isotropic and spherically symmetric stellar systems\footnote{\textit{Phaseflow} evolves the 1-D Fokker-Planck equation in the energy space, thus assuming the angular momentum distribution readjusts instantaneously to its relaxed,  equilibrium profile at each step. This approximation may not be always optimal \citep[][]{Broggi2022} but it renders the runs computationally much cheaper. More in general, additional physical processes as star formation (together with the shuffling in the galactic orbits induced by supernova kicks, \citealt{Bortolas2017, Bortolas2019}), the presence of a stellar disc in spite of a spherical distribution of stars \citep[][]{Souza-Lima2020}, as well as the presence of a central binary MBH \citep{Lezhnin2016, Li2017,Bortolas2018} may affect the rate estimates; these complex processes cannot be taken into account with our approach and require a more sophisticated modelling.}; it  solves the coupled system of Poisson and orbit-averaged
Fokker–Planck one-dimensional equations (depending only on the energy or -- to be more precise -- on the phase volume of states within a given energy) for the  potential, the density, and the distribution function. It allows to obtain the time evolution of the system as a result of two-body relaxation and stellar accretion by the MBH. This integrator has been successfully employed to track the rates of TDEs \citep[][]{Bortolas2022} and EMRIs \citep[][]{Pestoni2021}, systems dominated by primordial black holes \citep[][]{Zhu2018,Stegmann2020}, and nuclear stellar clusters \citep[][]{Generozov2018, Emami2020}.
For more details on the implementation see \citet{Vasiliev2017}. 
Our strategy consists of initialising the stellar systems and their central MBH as described below, and using \textit{Phaseflow} to extract the relevant quantities; for simplicity, we limit our analysis to a monochromatic population of 1~$\msun{}$ stars. We set the Coulomb logarithm that tunes the effects of two-body relaxation equal to $\ln\Lambda=\ln(0.4M_\bullet/\msun{})$ (\citealt{Spitzer1971}).  We stress here that for the purpose of the present paper, we do not evolve the stellar  systems through time; instead, our estimated event rates are  associated to the profiles chosen as initial conditions.\footnote{%
More specifically, \textit{PhaseFlow} provides us with the diffusion coefficients, orbital time, angular momentum associated to a circular orbit, distribution function; all those quantities are obtained as a function of $E$. With those, we can compute the PTDE and TTDE rates as a function of each ($E, \mathcal{R}$)  through Eq. \ref{eq:R}--\ref{eq:Rlc}, \ref{eq:alpha}--\ref{eq:delta_rep}.
}

In what follows, we  assume that PTDEs can occur up to a radius that strips a variable fraction of the initial stellar mass -- this choice defines the size of our loss cone. Our fiducial model assumes PTDEs can occur up to a radius that strips 3 per cent of the original stellar mass. This choice is somewhat arbitrary; in fact, the loss cone should have a radius that reaches the point at which the Cohn–Kulsrud boundary condition is applicable. Knowing exactly where to place this boundary would require a much more detailed knowledge of the PTDE process for each value of $\Delta m/m$, together with knowing the exact limit at which we can categorize the close passage as a PTDE instead of a mild tidal deformation of the star that could still be treated as a non-interacting object; this is to our knowledge still an open issue. Our choice of 3 per cent  happens to reflect the amount of accreted mass inferred from the light curves of ``faint'' TDE candidates, possibly indicating the minimum stripped mass for ``observable'' events 
\citep[e.g.,][]{Blagorodnova+2017, Nicholl+2020, Malyali+2023,Malyali+2023b}. 
Although we set $\Delta m/m=$ 3 per cent to be our fiducial value, we explore the dependence of our rate estimate on the values of $\Delta m/m$.

\section{Milky Way model}\label{sec:MW}

We first focus on a system that resembles the Milky Way nucleus. We initialise our system as in \citet{Bortolas2022}, with a \citet{Sersic1968} model for the galactic bulge and the central nuclear star cluster. We set the bulge stellar mass to $9.1\times 10^9 \msun$ \citep{Licquia2015}, its effective radius to $1.04$ kpc, and its Sersic index $ 1.3$ \citep{Davis2019}. On top of it, we set up a model for the nuclear star cluster based on \citet{Pfister2020, Schodel2017}: the cluster   effective radius is 6 pc, its Sersic index is equal to 2, and  its total mass is $ 4\times 10^7 \msun$. The central MBH 
 has $4\times10^6 \msun$ \citep{Gravity2020}. The Milky Way desity profile is shown with a black line in Fig.~\ref{fig:ic_profiles}.

 Our analysis revolves around estimating the rate of PTDE and TTDEs with our novel approach presented from Sec.~\ref{sec:ptde_theory}, for which the  loss cone is defined through the largest PTDE radius that guarantees an event to occur; the obtained rate of PTDE and TTDE within this model is compared with the standard estimate of TTDEs that  instead defines the loss cone through the radius at which a TTDE can occur.

\subsection{Event rates}

\begin{figure}
\includegraphics[ width=0.45\textwidth]{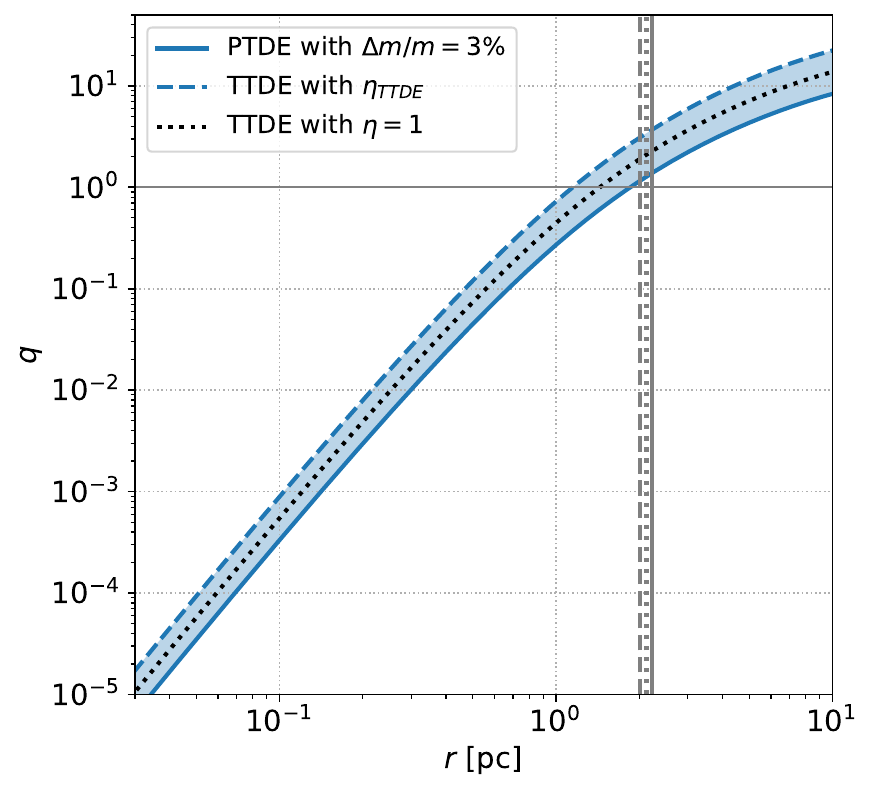}
    \caption{ Value of the loss cone filling factor $q$ as a function of the separation from the centre of the system for a Milky Way model (Sec.~\ref{sec:MW}). The value of $q$ is computed assuming a disruption radius equal to the PTDE radius that strips 3 per cent of the  original stellar mass (solid blue line, corresponding to our fiducial assumption througout this paper) and to the TTDE radius (dashed blue line). For reference we also show the value of $q$ for TTDEs assuming $\eta=1$ in Eq.~\ref{eq:rt} (black dotted line). The gray vertical lines mark the radius at which the disruption rate is maximal in the three different cases (line styles match the ones in the legend).}
    \label{fig:Milky_Way_q}
\end{figure}

 Fig.~\ref{fig:Milky_Way_q} shows the value of the loss cone filling factor $q$ as a function of radius for different values of  $r_t$ that define the loss cone. Since $q\propto 1/r_t$ (Eq.~\ref{eq:q}), the value of $q$ associated with PTDEs is systematically lower than the one computed using  $r_t$ for TTDEs, by an amount that is roughly equal to $\kappa$ ($\approx 2.68$ in this model; Sec.~\ref{sec:ptde_theory}). This implies that allowing for PTDEs to occur automatically implies a larger reservoir of stars are found in the empty loss cone.\footnote{This is very intuitive as a larger $\mathcal{R_{\rm lc}}$ implies two-body relaxation needs more time to change $\mathcal{R}$ by an amount of the order of $\mathcal{R_{\rm lc}}$ itself.} For reference, we find that the value of $q$ at which the event rate is maximum is 1.375.

\begin{figure}
\includegraphics[ width=0.45\textwidth]{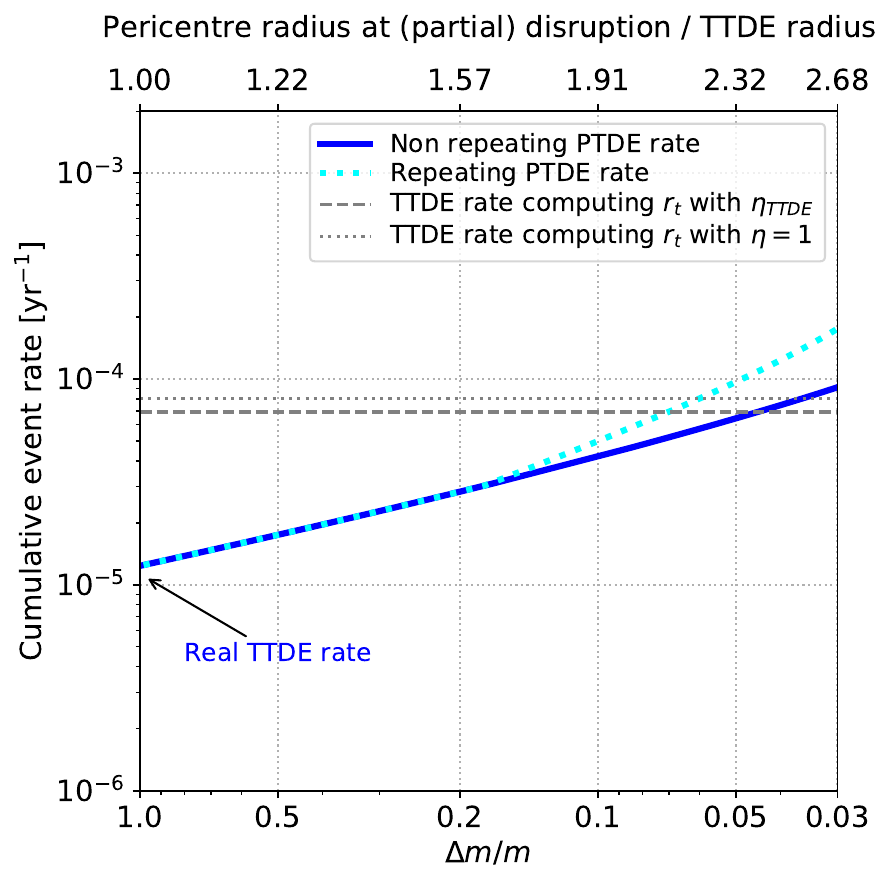}
    \caption{Cumulative TTDE+PTDE rates as a function of the relative stellar mass stripped in the event. The blue solid curve does not allow for repeating PTDEs, while the cyan dotted one assumes a star can undergo multiple PTDEs. The left-most value of the blue curve corresponds to the effective TTDE rate in a system in which PTDEs are allowed to occur. For comparison, the horizontal grey lines mark the TTDE rate one would have estimated neglecting PTDEs, and using either $\eta_{\rm TTDE}$ (dashed) or $\eta=1$ (dotted) in Eq.~\ref{eq:rt}. For reference, the top axis shows the value of the PTDE radius with respect to the TTDE radius (computed with $\eta_{\rm TTDE}$) for the different values of $\Delta m/m$. Remarkably, standard estimates of TTDE rates (grey horizontal lines) can easily overestimate the real TTDE rate by almost one order of magnitude, implying that properly accounting for PTDEs is crucial to obtain reliable event rates, especially in systems (as the present one) in which a significant fraction of the total flux comes from the empty loss cone.} 
    \label{fig:Milky_Way_PTDE}
\end{figure}

Fig.~\ref{fig:Milky_Way_PTDE} displays the cumulative amount of PTDEs+TTDEs (9.1$\times10^{-5}$ yr$^{-1}$ in total) as a function of the fraction of stellar mass stripped in the event. The left-most  point in the curve is in fact the TTDE rate in our model (1.2$\times10^{-5}$ yr$^{-1}$, indicated by an arrow with ``real TTDE rate'' at $\Delta m/m=1$). There are two important features that highlight the importance of PTDEs in this figure.  First of all, the overall rate of PTDEs resulting in a mass loss larger than 3 percent of the original stellar mass  can be greater than the amount of TTDEs happening in the same system by up to one order of magnitude. 
If a star can undergo repeating PTDEs, then the overall rate of PTDEs+TTDEs  becomes further enhanced by a factor $\approx 2$. Second, the plot shows that the TTDE rate computed assuming the standard loss cone size without any consideration of PTDEs (dashed or dotted grey lines) 
results in a significant overestimate of the TTDE rate, by nearly one order of magnitude. This result can be understood in the following terms: stars coming from the full loss cone have a probability to be disrupted at each value of $r<r_t$ which is nearly flat; since the radius at which TTDEs can occur is $\kappa \approx2.68$ times smaller than the loss cone radius adopted, the TTDEs coming from the full loss cone should be $\approx 0.37$ times the PTDEs coming from the same regime. Stars coming from the empty loss cone have an even more dramatic impact in the TTDE rate: since they enter the loss cone with small steps in angular momentum, they are way more likely to undergo an event at $r$ very close to the largest $r_t$ that allows PTDEs to occur. This implies that virtually all stars coming from the empty loss cone can only undergo PTDEs, while if we were not accounting for PTDEs in our model, these stars would  eventually be disrupted in a TTDE (see Fig.~\ref{fig:cartoon} for a visual representation of this concept). 
More quantitatively, we can estimate the fraction of TTDEs over PTDEs via Eq.~\ref{eq:delta_ttde} computing $\alpha$ ($\approx2.34$) at the value of $E$ at which the overall TDE flux is maximum; we get $\delta_{\rm TTDE}\approx$ 16 per cent, comparable to the actual value of 13.4 per cent.  

\subsection{Choice of the loss cone outer boundary for PTDEs}

\begin{figure}
\includegraphics[ width=0.45\textwidth]{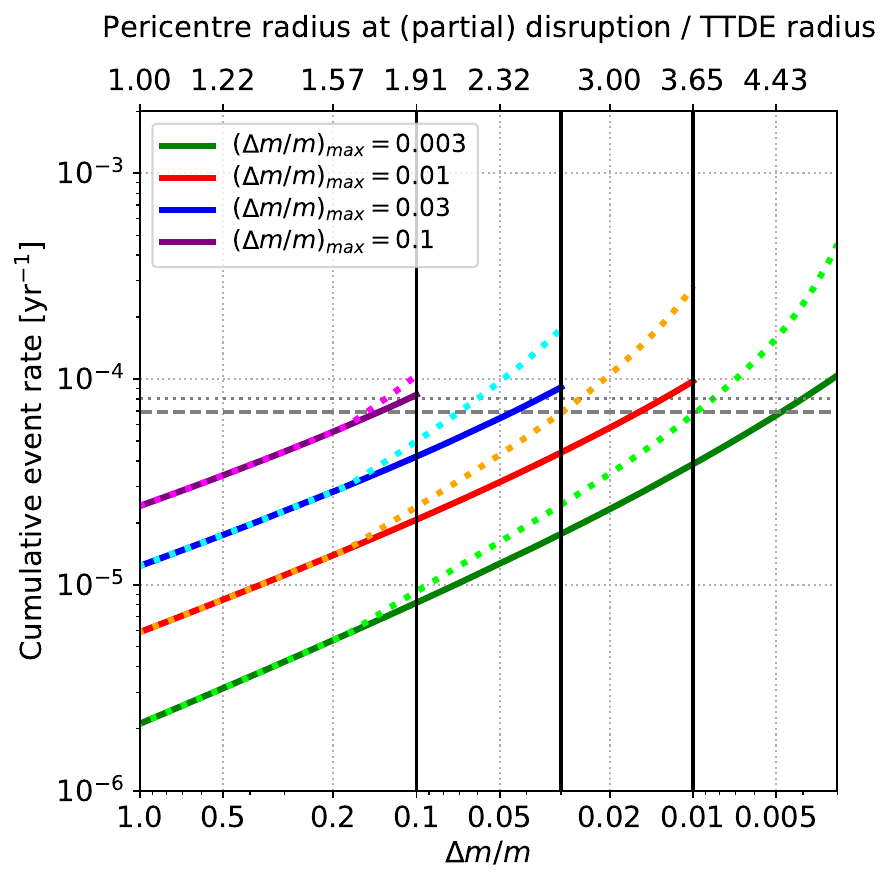}
    \caption{Cumulative PTDE+TTDE rate as a function of the relative stellar mass stripped in the event (i.e., same as Fig.~\ref{fig:Milky_Way_PTDE}, note the different x-axis range) for different choice of the minimum $\Delta m/m$ (or analogously, the $r_t$ defining the loss cone) for which PTDEs are allowed to occur and are properly described within our framework. In particular, here the different lines show the non-repeating (solid) or repeating (dotted) PTDE rates for a minimum $\Delta m/m = 10\%$ ($\kappa = 1.91$, purple), $\Delta m/m = 3\%$ ($\kappa = 2.68$, blue), $\Delta m/m = 1\%$ ($\kappa = 3.65$, red) and $\Delta m/m = 0.3\%$ ($\kappa = 5.12$, green). Note that the total event rate in the non-repeating assumption (the right-most point of each curve) yields a very similar result in all cases ($\approx 10^{-4}$ yr$^{-1}$); on the other hand, the amount of actual TTDEs (i.e. the left-most point in each curve) varies non negligibly depending on the choice of the minimum allowed $\Delta m/m$. These results are expected for a system in which a significant amount of the total TDE flux comes from the empty loss cone.}
    \label{fig:change_rlc}
\end{figure}

An important caveat to keep in mind is the fact that the choice of the last available radius at which we allow PTDEs to occur (i.e. the radius that defines our loss cone) has a non-negligible impact on our estimate of the TTDE rates. Fig.~\ref{fig:change_rlc} shows different trends in the cumulative PTDE rate depending on the threshold $\Delta m/m$ above which we allow PTDEs to occur. Although in this specific model the total amount of (non-repeating) events per year has only a mild dependence on this choice (as it remains always close to $10^{-4}$ yr$^{-1}$, as expected in a system dominated by the empty loss-cone regime), the amount of TTDEs in each model varies by nearly an order of magnitude if we change the threshold for PTDEs to occur from $\Delta m/m=0.1$ to $3\times10^{-3}$; as expected, a decreasing limiting $\Delta m/m$ (thus an increasing size of the loss cone) results in a systematically lower TTDE rate, as most events happen at the largest available radius and a systematically smaller fraction can reach the limiting distance below which TTDEs can occur. Fig.~\ref{fig:change_rlc} also shows that the overall amount of repeating PTDEs gets larger as the limiting $\Delta m/m$ is lowered: in fact, in this regime the number of orbits a star needs to undergo in order to be totally consumed through a series of repeating PTDEs gets larger and larger; furthermore, the loss cone gets systematically emptier. 

Although this is an intrinsic limitation of our current modelling, it should be noticed that the rate strongly depends on the lower limit of $\Delta{m}/m$ only in dense stellar models, in which most of the disrupted stars come from the empty loss cone regime. In the next section we will show that the dependence  of the TTDE on the choice of the limiting $\Delta m/m$  is not as dramatic for systems dominated by the full loss cone.

\section{A broader set of models}\label{sec:multimodel}

\begin{figure}
\includegraphics[ width=0.45\textwidth]{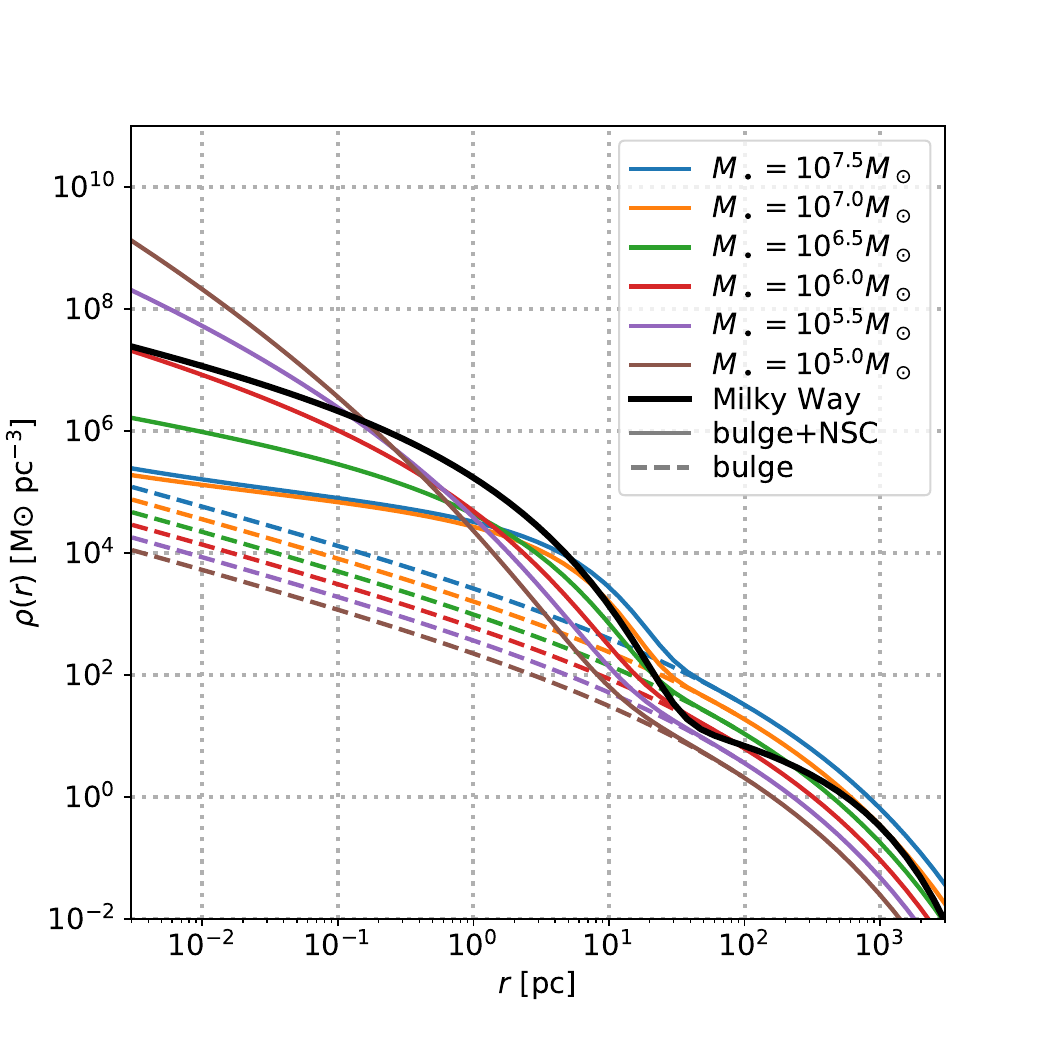}
    \caption{Initial density profiles chosen for our systems, depending on the mass $M_\bullet$ of the central MBH; the associated properties are illustrated in Tab.~\ref{tab:IC}. 
 We also show the Milky Way model adopted in Sec.~\ref{sec:MW} with a black line. The dashed lines show the profiles accounting for the stellar bulge only, while solid lines show the profiles that additionally feature a nuclear stellar cluster. }
    \label{fig:ic_profiles}
\end{figure}

We now use the Fokker Planck integrator \textit{Phaseflow} to estimate the rate of PTDEs and TTDEs about MBHs with different masses $M_\bullet$. In particular, we explore the following cases: $M_\bullet = \{10^{7.5},10^7, 10^{6.5}, 10^6, 10^{5.5}, 10^5\} \msun{}$.
We build the host stellar system about the MBHs based on characteristic scaling relations between the MBH mass and its host.
For each MBH mass, we explore two possibile host stellar systems: (i) featuring both a stellar bulge and a nuclear stellar cluster; (ii) featuring only a stellar bulge.
In the bulge-only case, the stellar bulge density profile is chosen as a deprojected \citet{Sersic1968} profile with total stellar mass (\citealt{Greene2020}, see their tab.~9, case dubbed \textit{all, limits})
\begin{equation}
M_b/\msun{} = \left(3\times 10^{10} \msun{}\right) 10^{[{\log(M_\bullet/\msun{})-7.43}]/{1.61}};
\end{equation}
the \citeauthor{Sersic1968} effective radius is modelled via 
\begin{equation}
    R_{e, b}/{\rm pc} = 10^{0.14\log(M_b/\msun{})+1.79}
\end{equation}
(\citealt{Shen2003}, case for low-mass late-type galaxies, eq.~32); for simplicity, we keep the  \citeauthor{Sersic1968} index equal to  $n=2.5$ for all MBH masses. 
When the nuclear cluster is  present, we model its properties based on the relations presented in \citet{Pechetti2020}. Specifically, the cluster follows a \citet{Sersic1968} density profile with total mass 
\begin{equation}
    M_n/\msun{} = 10^{6.308+0.94\, \log(M_b/10^9\msun{})},
\end{equation}
effective radius
\begin{equation}
    R_{e, n}/{\rm pc} = 10^{0.53+0.29\, \log(M_b/10^9\msun{})},
\end{equation}
and \citeauthor{Sersic1968} index
\begin{equation}
    n_{n} = 10^{-0.66\log(M_n/10^6\msun{})+0.95}.     
\end{equation}
Since no isotropic model with inner density slope shallower than $\rho\propto r^{-1/2}$ can be built about an MBH, we set the values of $n_{n}<1$ equal to $n_{n}=1$. 
Fig.~\ref{fig:ic_profiles} displays the  density profiles as obtained from the previous equations, while 
Tab.~\ref{tab:IC} reports the values of the quantities adopted for initialising each system.

\begin{table}
	\centering
	\caption{Parameters adopted for initializing the stellar system associated to each MBH mass (first column). The following columns respectively report the bulge total mass and effective radius; in the case of runs accounting for a nuclear stellar cluster, the last three columns respectively show the cluster total mass,  effective radius and Sersic index.}
	\label{tab:IC}
	\begin{tabular}{rccccc} 
		\hline
		$M_\bullet/\msun{}$  & $M_b/\msun{}$  & $R_{e,b}/$pc & $M_n/\msun{}$ & $R_{e,n}/$pc & $n_n$\\
		\hline
        $10^{7.5}$ &  $3.316\times 10^{10}$ & 1,832 & $5.462\times 10^7$ &  9.353 & 1.00 \\
        $10^{7}$   &  $1.622\times 10^{10}$ & 1,657 & $2.789\times 10^7$ & 7.602  & 1.00 \\
        $10^{6.5}$ &  $7.934\times 10^{9} $ & 1,499 & $1.424\times 10^7$ & 6.178  & 1.54 \\
        $10^{6}$   &  $3.881\times 10^{9} $ & 1,357 & $7.271\times 10^6$ & 5.021  & 2.41 \\
        $10^{5.5}$ &  $1.898\times 10^{9} $ & 1,227 & $3.712\times 10^6$ & 4.081  & 3.75 \\
        $10^{5}$   &  $9.285\times 10^{8} $ & 1,110 & $1.896\times 10^6$ & 3.316  & 5.84 \\
		\hline
	\end{tabular}
\end{table}

\begin{figure*}
\includegraphics[ width=0.49\textwidth]{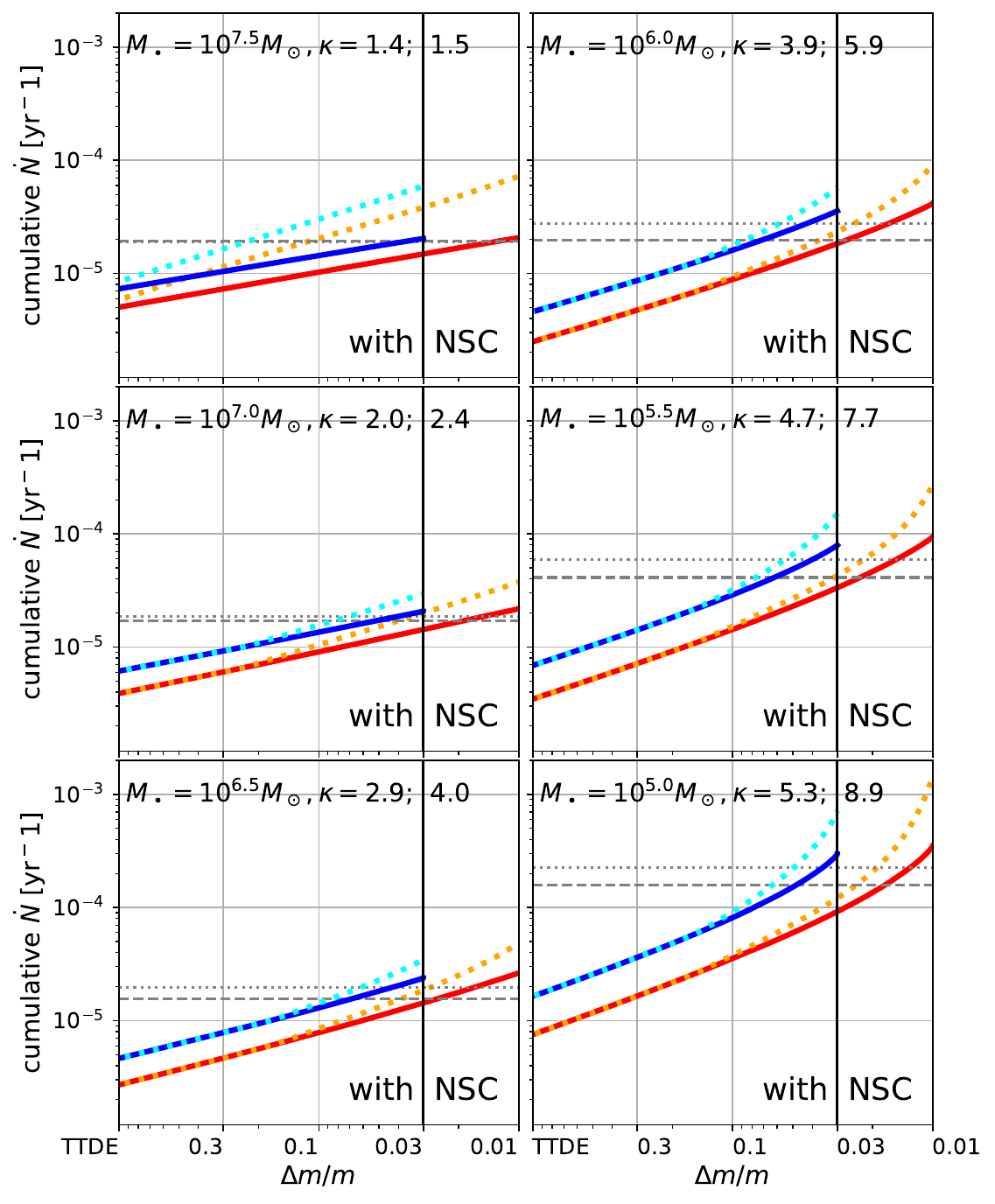}
\includegraphics[ width=0.49\textwidth]{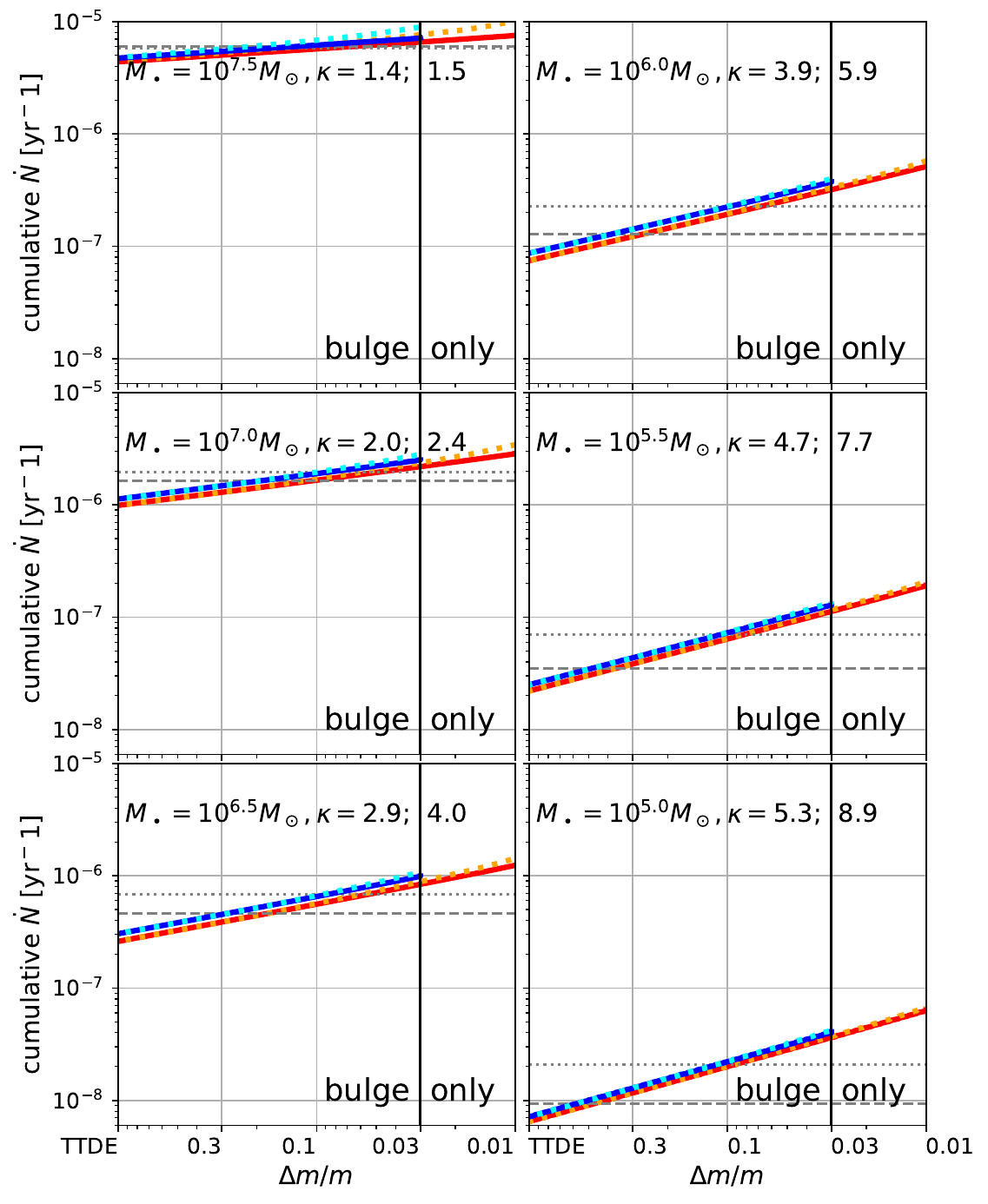}
    \caption{ The panels shows the cumulative PTDE+TTDE rate as a function of the relative stellar mass stripped in the event (as Fig.~\ref{fig:Milky_Way_PTDE}) for two different choice of the minimum $\Delta m/m$ for which PTDEs can occur: the different lines show the non-repeating (solid) or repeating (dotted) PTDE rates for a minimum $\Delta m/m = 3\%$ (blue and cyan) and $\Delta m/m = 1\%$ (red and orange); the associated values of $\kappa$ are shown in the top right of the panels. For comparison, in each panel the horizontal grey lines mark the TTDE rate one would have estimated neglecting PTDEs, and using either $\eta_{\rm TTDE}$ (dashed) or $\eta=1$ (dotted) in Eq.~\ref{eq:rt}.  The six left panels refer to runs with the nuclear star cluster, while the six panels on the right show the bulge-only runs; note that the y-axis is different in the runs with and without the nuclear stellar cluster, even if its span in dex is the same.
    The MBH mass varies in each panel as written in the top-left corner. Tables~\ref{tab:rates0.03}, \ref{tab:rates0.01} provide a more quantitative perspective on the results presented here. }
    \label{fig:rates_multimodel}
\end{figure*}

\begin{table*}
	\centering
	\caption{This table refers to runs for which the maximum value of $\Delta m/m$ for which PTDEs can occur is 3 per cent. For each model (column 1), we show the associated value of $\kappa$ for the given MBH mass (column 2), the value of $q$ at the energy for which the TDE flux is maximum (column 3), the PTDE+TTDE rate in units of yr$^{-1}$ (column 4),  the TTDE rate in units of yr$^{-1}$ (column 5), the fraction of events (PTDEs+TTDEs) that are TTDEs (column 6), the rate of  PTDE+TTDEs assuming that the former can be repeated -- in units of  yr$^{-1}$ (column 7), the fraction of TTDEs that are not visible as they directly end up inside the MBH  horizon (with radius $2GM_\bullet/c^2$, column 8), the TTDE rate computed without accounting for PTDEs and assuming $\eta_{\rm TTDE}$ (column 9) or $\eta=1$ (column 10) in Eq.~\ref{eq:rt} divided by the  TTDE rates estimated in our models that also account for PTDEs.}
	\label{tab:rates0.03}
	\begin{tabular}{rccccccccc}
		\hline
		Model  & $\kappa$ & $q$ & P+TTDE  & TTDE & TTDE/P+TTDE & P+TTDE rep. & horizon/TTDE &  $\eta_{\rm TTDE}$ overest. &  $\eta=1$ overest. \\
		\hline
            $10^{7.5}\msun{}$ NSC &  1.382 & 0.23 & $2.036 \times 10^{-5}$ & $7.314 \times 10^{-6}$ & 0.3592 & $5.928 \times 10^{-5}$ & 0.08068 &  2.636 &  2.611\\
$10^{7.0}\msun{}$ NSC &  1.978 & 0.73 & $2.069 \times 10^{-5}$ & $6.133 \times 10^{-6}$ & 0.2964 & $2.953 \times 10^{-5}$ & 0.1037 &  2.763 &   3.02\\
$10^{6.5}\msun{}$ NSC &  2.881 & 0.63 & $2.375 \times 10^{-5}$ & $4.635 \times 10^{-6}$ & 0.1951 & $3.439 \times 10^{-5}$ & 0.08125 &  3.362 &   4.23\\
$10^{6.0}\msun{}$ NSC &  3.885 & 0.37 & $3.57 \times 10^{-5}$ & $4.59 \times 10^{-6}$ & 0.1286 & $5.751 \times 10^{-5}$ & 0.0459 &    4.3 &  5.992\\
$10^{5.5}\msun{}$ NSC &  4.721 & 0.09 & $7.947 \times 10^{-5}$ & $6.884 \times 10^{-6}$ & 0.08662 & $1.515 \times 10^{-4}$ & 0.02197 &  5.986 &  8.669\\
$10^{5.0}\msun{}$ NSC &  5.271 & $\approx0$ & $2.959 \times 10^{-4}$ & $1.641 \times 10^{-5}$ & 0.05544 & $7.127 \times 10^{-4}$ & 0.009255 &  9.608 &  13.73\\
\hline
$10^{7.5}\msun{}$ bulge &  1.382 & 1.48 & $7.145 \times 10^{-6}$ & $4.736 \times 10^{-6}$ & 0.6629 & $8.97 \times 10^{-6}$ & 0.3423 &  1.264 &  1.227\\
$10^{7.0}\msun{}$ bulge &  1.978 & 2.23 & $2.515 \times 10^{-6}$ & $1.127 \times 10^{-6}$ & 0.4481 & $2.818 \times 10^{-6}$ & 0.2248 &  1.447 &  1.739\\
$10^{6.5}\msun{}$ bulge &  2.881 & 4.11 & $9.924 \times 10^{-7}$ & $3.047 \times 10^{-7}$ &  0.307 & $1.083 \times 10^{-6}$ & 0.1367 &  1.512 &  2.246\\
$10^{6.0}\msun{}$ bulge &  3.885 & 6.06 & $3.758 \times 10^{-7}$ & $8.672 \times 10^{-8}$ & 0.2308 & $4.029 \times 10^{-7}$ & 0.07699 &   1.47 &  2.611\\
$10^{5.5}\msun{}$ bulge &  4.721 & 9.41 & $1.297 \times 10^{-7}$ & $2.503 \times 10^{-8}$ & 0.1929 & $1.37 \times 10^{-7}$ & 0.04062 &  1.389 &  2.813\\
$10^{5.0}\msun{}$ bulge &  5.271 & 11.78 & $4.088 \times 10^{-8}$ & $7.161 \times 10^{-9}$ & 0.1752 & $4.271 \times 10^{-8}$ & 0.02045 &  1.313 &  2.906\\
        
		\hline
	\end{tabular}
\end{table*}

\subsection{Event rates}

Fig.~\ref{fig:rates_multimodel} shows the analogous of Fig.~\ref{fig:Milky_Way_PTDE} for the generalized models and for two different choices of the PTDE largest allowed radius (stripping either 3 or 1 per cent of the original stellar mass). Tab.~\ref{tab:rates0.03} reports the main results obtained through the modelling of the same system assuming the minimum $\Delta m/m$ for PTDEs to occur is 3 per cent; in the appendix, we show analogous tables for different choices of this minimum allowed $\Delta m/m$ (Tab.~\ref{tab:rates0.003},~\ref{tab:rates0.01},~\ref{tab:rates0.1}).

The event rates  are much larger in the runs featuring a nuclear star cluster compared with the bulge only runs, as expected; this is due to the larger stellar density of nucleated models, which results in more efficient relaxation. 
Assuming a limiting $\Delta m/m=0.03$, TTDE rates in nucleated galaxies are close to $10^{-5}$ yr$^{-1}$, while PTDE rates stay in the range $10^{-5}-10^{-4}$ yr$^{-1}$ and get close to $10^{-3}$ yr$^{-1}$ if we allow for repeating PTDEs. In bulge-only systems, the rates are significantly smaller and even accounting for repeating events, they never reach $10^{-5}$ yr$^{-1}$. These event rates are  bound to our specific choice of initial conditions.
An interesting feature of the selected models is that the nucleated systems tend to be dominated by the empty loss cone (Tab.~\ref{tab:rates0.03}  shows that  $q<1$);  the bulge-only systems are dominated by the full loss cone instead. This distinction allows us to analyse the rate of TTDEs over the total number of events in the two  regimes.  

Before going into the detailed estimates of the rates, it is important  to keep in mind that with our definition of the loss cone,  $q\propto 1/r_t\propto 1/\kappa$, so the loss-cone is emptier if $\kappa$ is larger (or equivalently, the limiting $\Delta m/m$ is smaller). $\kappa$ grows as the MBH mass gets smaller. Most literature works  (not accounting for PTDEs) find that $q$ generally increases as the MBH mass gets smaller \citep[e.g.][]{Wang2004, Stone2016}; this is bound to the fact that -- if one neglects PTDEs -- the loss cone size scales as $r_t\propto M_\bullet^{1/3}$. Yet, in our framework, $\kappa$ grows significantly as the MBH mass decreases, so that the drop in $r_t$ with the MBH mass is not as significant; as a consequence, the loss cone does not always get more full as the MBH mass decreases (see  in Tab.~\ref{tab:rates0.03}).\footnote{The dependence of $r_t$ on the MBH mass under different assumptions is shown in the appendix in Fig.~\ref{fig:radius_k}.}

Let us now focus on TTDE rates: they are almost always subdominant by a factor that ranges from a few to more than one order of magnitude with respect to PTDEs (even neglecting repeating ones). The fraction of TTDEs over the total number of events gets systematically  smaller as the MBH mass decreases; the same fraction is systematically larger in non-nucleated galaxies. Fig.~\ref{fig:rates_multimodel} and Tab.~\ref{tab:rates0.03} also compare the estimate of TTDEs obtained in our framework with the one obtained traditionally, using $\eta=1$ or $\eta_{\rm TTDE}$ to define the loss cone and neglecting PTDEs. We find that the traditional estimates of TTDEs are overestimated by a factor that gets larger for nucleated systems (as their loss cone is more empty),   smaller MBH masses (as their associated $\kappa$ is larger), and smaller limiting $\Delta m/m$ (again due to their larger $\kappa$).  

In the full loss cone systems (bulge only) the event rate scales linearly with the size of the loss cone itself; as a consequence, changing the size of the loss cone does not dramatically impact the TTDE rate, and even choosing a different limiting $\Delta m/m$ nearly yields the same results. This is even more true for systems characterized by larger $M_\bullet$, for which $\kappa$ is smaller.

Systems characterised by empty loss cones (nucleated ones) behave differently. Here most events occur as PTDEs, and TTDEs are systematically suppressed, as only the smaller set of stars belonging to the full loss cone can penetrate deeper near the MBH and result in a TTDE.  
Tab.~\ref{tab:rates0.03} shows that traditional rates of TTDEs are overestimated  by factors  that can be larger than an order of magnitude around small MBHs. Furthermore, in nucleated systems, the choice of the limiting $\Delta m/m$ has a non negligible impact on the final results.

Let us assume one wants to quantify what is the fraction of events that end up as TTDEs over the entirety of events (PTDEs \textit{plus} TTDEs. This fraction is trivially estimated through $1/\kappa$ in systems with a relatively full loss cone, as expected. To make the same estimate in systems with an empty loss cone, in principle one should multiply $1/\kappa$ by the fraction of stars that are found in the full loss cone.
Yet, we can use an empirical estimate by noticing that systems with an empty loss cone have the  fraction of TTDEs over total events which is  well approximated by $1/\kappa^2$ (within a factor $\sim$2).

Concerning repeating PTDEs, those are much more likely to occur in the empty loss cone: this is reflected in the fact that  the overall event rates get enhanced by up to a factor $\approx 10$ in the nucleated galaxies, while the enhancement is almost negligible in the bulge-only runs. 

Finally, Tab.~\ref{tab:rates0.03}  quantifies the fraction of TTDEs that end up directly in the MBH horizon and cannot yield an observable TTDE.  Those are generally a small fraction of the total  and do not impact our overall estimate of TTDEs, except around the largest considered MBHs and in systems dominated by the full loss cone, as expected.

\section{Summary and discussion}\label{sec:discussion}

In this paper, we investigated the rates of PTDEs (and how they compare with the rates of TTDEs) about MBHs of different masses and stellar systems with different properties (a Milky Way model, a set of systems with and without a nuclear stellar cluster).
The core choice of our modelling strategy is to define the loss cone through the largest radius at which PTDEs can occur -- instead of the largest radius at which TTDEs occur. We argue that this is a more sensible way to estimate both PTDE and TTDE rates, and we compare our estimate of TTDE rates with the one  obtained  defining the loss cone  with the traditional estimate of the tidal radius 
 $r_t\approx(M_\bullet/m_\star)^{1/3}R_\star$.

Below we summarize our key results.
\begin{enumerate}
    
    \item PTDEs are generally more abundant than TTDEs by a factor of a few to a few tens;  the PTDE rate is further enhanced when we assume the same star can undergo multiple PTDEs.
    \item Centrally concentrated systems (e.g. hosting a nuclear stellar cluster) tend to have an emptier loss cone. Here our new definition of loss cone significantly reduces (by a factor of a few to a few tens) the overall TTDE rates compared with estimates in which the loss cone is traditionally defined. In empty loss cone  systems the same star is more likely to undergo a series of repeated PTDEs, so that the overall PTDE rate can get larger by one order of magnitude once accounting for this.
    \item Bulge-only systems tend to have a fuller loss cone. Here the traditional computation of TTDE rates yield results that are only mildly larger than the rates we obtained in this work. In these systems, repeating PTDEs are also less likely to occur.
    \item The choice of the limiting stripped stellar mass $\Delta m/m$ above which PTDEs can occur defines our loss cone: this choice impacts our results in systems characterised by an empty loss cone (nucleated galaxies), while it has a minor impact in bulge-only systems. Our fiducial model sets the limiting $\Delta m/m$ to 3 per cent.
\end{enumerate}

An interesting point worth making is that our analysis brings to an important reduction in the overall number of TTDEs in many systems, compared with traditional TTDE rate estimates. The reduction is more relevant in galaxies with a nuclear star cluster; since these systems have much larger event rates than bulge-only systems, they are expected to host most of the observed events.
This consideration is important as the theoretical rates of TTDEs may be 
larger than the  observed ones by nearly an order of magnitude \citep[e.g.][]{French2020,Sazonov+2021,Lin2022}.  Our novel treatment can help reconciling the observed and modelled rates of TTDEs, given that PTDEs are much dimmer thus likely much harder to detect. In addition,   \citet[]{Mockler2019} used the light-curve of observed TDEs to infer the mass of the destroying MBH and (although with higher uncertainty) that of the star: most of the destroyed masses are low ($\sim 0.1 \msun{}$, see their tab.~2) supporting the idea that a PTDE has occurred and only a minor fraction of the total stellar mass has been accreted by the MBH.

A further relevant consideration is related to the the TDE luminosity function, which has been constrained by several authors in the last years \citep{vanVelzen2018,Sazonov+2021,Lin2022}, and whose shape and normalisation necessarily has an  impact on the   rate of TDEs per galaxy as derived from observations. PTDEs are intrinsically dimmer, yet when detected, it may be challenging to distinguish them from TTDEs based only on the decline of their light-curve. As a consequence, unless a repeating event is observed, PTDEs and TTDEs may  be ascribed to the same parent population. Given these considerations,  it is crucial to account for PTDEs both when theoretically estimating the event rates of TDEs, and when constraining the luminosity function of these sources. Future work in this direction will be crucial to use TDE flares to constrain the population and occupation fraction of MBHs.

Our results are obtained under a set of assumptions. As mentioned above (Sec.~\ref{sec:ptde_rate}), \textit{repeating} PTDEs can only occur if the stellar orbit, mass and radius do not dramatically vary during each PTDE; in this context, our \textit{non-repeating} PTDE estimate is the most conservative. Assessing how many PTDEs can be experienced by the same star in detail depending on the MBH and stellar properties would be fundamental to better constrain this aspect. In this same context, it is possible that stars undergoing a PTDE are  deflected onto an orbit that results in the entire disruption of the leftover remnant after the first PTDE; if the light-curves associated with these latter events are indistinguishable from those of TTDEs, it means the  fraction of TTDEs we estimate here can be lower than the one of \textit{observed} events dubbed as {TTDEs}. 
A second important limitation is our modelling of the system surrounding the central MBH. Although our implementation is based on observational constraints, it is important to stress that  the scaling between the MBH mass and the host properties are likely poor towards low MBH masses, where less MBH mass measurements are available \citep[][]{Greene2020}. It is also worth stressing that our rate estimates are affected by the choice of the smallest amount of stripped mass -- or equivalently the largest radius -- at 
which PTDEs occur. Varying this value affects our results, highlighting the importance of assessing this aspect for properly estimating PTDE and TTDE rates. Further limitations include the fact that we did not evolve our model through time and we only accounted for a monochromatic stellar mass function. \citet{Bortolas2022} showed that, if a non-monochromatic stellar mass function is considered, the evolution of the TTDE rate can be significantly different from that for a monochromatic mass function, such as a strong TDE burst over the first 0.1 - 1 Gyr in post-starburst galaxies with a complete mass function. This naturally indicates a possible modification of PTDE rates when a realistic mass function is taken into account. Furthermore, evolved stars as giants are intrinsically more prone to PTDEs due to their stellar structure \citep{MacLeod2012}.  Considering these aspects will be the focus of our future work.

In this study, we focused on comparing the rate of theoretical PTDEs to that of theoretical TTDEs. Although our results will provide a useful  guideline for utilizing observational data, accurately estimating observable event rates is imperative. PTDEs may create light curves distinctive from those of TTDEs. A smaller debris mass produced in each PTDE translates to a lower total radiated energy than that for TTDEs. The shape of light curves for PTDEs remains still quite uncertain. Based on the debris' properties near the first pericenter passage found in previous numerical works \citep[e.g.,][]{Goicovic+2019, Ryu+2020c}, one might expect that the peak luminosity of PTDEs is lower and the post-peak luminosity decreases more rapidly than that for TTDEs. In fact, some candidates have shown hints of these features \citep{Hinkle+2020}. On the other hand, repeating PTDEs, in principle, can result in quasi-periodic light curves with peak-to-peak separations $\sim$ the orbital period of remnants. But if peak-to-peak separations are too long \citep{Ryu+2020c} compared to the span of transient surveys, it would be difficult to observe such quasi-periodic light curves. Estimating the rate of observable PTDEs with all the consideration of their light curves including those above will be the focus of a future study.

To conclude, our work implies PTDEs should be accounted for when assessing the expected TDE rates accessible to forthcoming facilities. Although the smaller mass available in each PTDE results in a less luminous electromagnetic emission, PTDEs may be observed when they are generated in nearby galaxies. Furthermore, their occurrence may impact the relative mass growth of the central MBH. Finally, not accounting for PTDEs can result in significant overestimates of the TTDE rates.

\section*{Acknowledgements}
We warmly thank the anonymous referee for their comments and suggestions.  EB and AS acknowledge support from the European Research Council (ERC) under the European Union's Horizon 2020 research and innovation program ERC-2018-CoG
under grant agreement N.~818691 (B~Massive). EB  acknowledges support from the European Consortium for Astroparticle Theory in the form of an Exchange Travel Grant, and the
European Union’s Horizon 2020 Programme under the AHEAD2020 project (grant
agreement n.~871158).

\section*{Data Availability Statement}
The data underlying this article will be shared on reasonable request to the corresponding author.




\bibliography{bibliography} 

\appendix 
\section{Additional tables and figures}

\begin{figure}
\includegraphics[ width=0.45\textwidth]{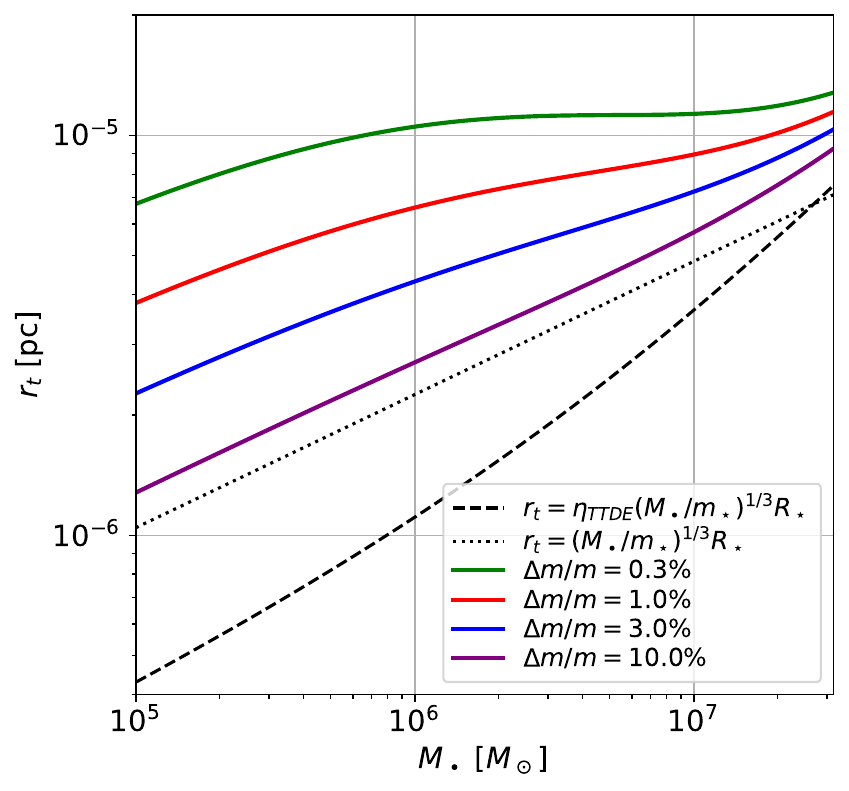}
    \caption{Tidal disruption radius $r_t$ as a function of the MBH mass. The dotted gray line shows the traditional definition of $r_t$ accounting only for TDEs and assuming $\eta=1$; the dashed grey line also accounts only for TTDEs but adopts Eq.~\ref{eq:eta_ftde} to define TTDEs. The colored lines show the value of $r_t$ if one defines it as the radius at which PTDEs can start occurring, as we do in this paper; different colors refer to the minimum allowed fraction of stellar mass above which PTDEs can occur.}
    \label{fig:radius_k}
\end{figure}

\begin{table*}
	\centering
	\caption{Same as Tab.~\ref{tab:rates0.03} assuming the maximum value of $\Delta m/m$ at which PTDEs can occur is 0.3 per cent. }
	\label{tab:rates0.003}
	\begin{tabular}{rccccccccc}
		\hline
		Model  & $\kappa$ & $q$ & P+TTDE  & TTDE & TTDE/P+TTDE & P+TTDE rep. & horizon/TTDE &  $\eta_{\rm TTDE}$ overest. &  $\eta=1$ overest. \\
		\hline
        $10^{7.5}\msun{}$ NSC &  1.709 & 0.19 & $2.106 \times 10^{-5}$ & $3.221 \times 10^{-6}$ &  0.153 & $8.265 \times 10^{-5}$ &  0.119 &  5.985 &  5.929\\
$10^{7.0}\msun{}$ NSC &  3.094 & 0.47 & $2.295 \times 10^{-5}$ & $2.089 \times 10^{-6}$ & 0.09103 & $4.67 \times 10^{-5}$ & 0.06749 &   8.11 &  8.866\\
$10^{6.5}\msun{}$ NSC &  5.772 & 0.36 & $2.911 \times 10^{-5}$ & $1.29 \times 10^{-6}$ & 0.04431 & $7.016 \times 10^{-5}$ & 0.05782 &  12.08 &   15.2\\
$10^{6.0}\msun{}$ NSC &   9.47 & 0.15 & $4.757 \times 10^{-5}$ & $1.089 \times 10^{-6}$ & 0.02289 & 0.0001579 & 0.03566 &  18.13 &  25.26\\
$10^{5.5}\msun{}$ NSC &  13.08 & $\approx 0$ & 0.0001103 & $1.41 \times 10^{-6}$ & 0.01278 & 0.0005669 & 0.01807 &  29.23 &  42.32\\
$10^{5.0}\msun{}$ NSC &   15.7 & $\approx 0$ & 0.000416 & $2.814 \times 10^{-6}$ & 0.006764 & 0.003933 & 0.007999 &  56.01 &  80.02\\
\hline
$10^{7.5}\msun{}$ bulge &  1.709 & 1.20 & $7.975 \times 10^{-6}$ & $4.026 \times 10^{-6}$ & 0.5048 & $1.178 \times 10^{-5}$ & 0.3417 &  1.487 &  1.443\\
$10^{7.0}\msun{}$ bulge &  3.094 & 1.65 & $3.257 \times 10^{-6}$ & $8.465 \times 10^{-7}$ & 0.2599 & $4.269 \times 10^{-6}$ & 0.2212 &  1.926 &  2.314\\
$10^{6.5}\msun{}$ bulge &  5.772 & 2.41 & $1.562 \times 10^{-6}$ & $2.136 \times 10^{-7}$ & 0.1368 & $1.94 \times 10^{-6}$ &  0.133 &  2.157 &  3.203\\
$10^{6.0}\msun{}$ bulge &   9.47 & 3.57 & $7.185 \times 10^{-7}$ & $6.126 \times 10^{-8}$ & 0.08525 & $8.518 \times 10^{-7}$ & 0.07468 &  2.081 &  3.696\\
$10^{5.5}\msun{}$ bulge &  13.08 & 6.17 & $2.913 \times 10^{-7}$ & $1.868 \times 10^{-8}$ & 0.06414 & $3.319 \times 10^{-7}$ & 0.03962 &  1.861 &  3.768\\
$10^{5.0}\msun{}$ bulge &   15.7 & 7.72 & $1.029 \times 10^{-7}$ & $5.708 \times 10^{-9}$ & 0.05549 & $1.138 \times 10^{-7}$ & 0.02007 &  1.647 &  3.645\\

		\hline
	\end{tabular}
\end{table*}

\begin{table*}
	\centering
	\caption{Same as Tab.~\ref{tab:rates0.03} assuming the maximum value of $\Delta m/m$ at which PTDEs can occur is 1 per cent. }
	\label{tab:rates0.01}
	\begin{tabular}{rccccccccc}
		\hline
		Model  & $\kappa$ & $q$ & P+TTDE  & TTDE & TTDE/P+TTDE & P+TTDE rep. & horizon/TTDE &  $\eta_{\rm TTDE}$ overest. &  $\eta=1$ overest. \\
		\hline
        $10^{7.5}\msun{}$ NSC &  1.529 & 0.21 & $2.069 \times 10^{-5}$ & $5.031 \times 10^{-6}$ & 0.2431 & $7.215 \times 10^{-5}$ & 0.09229 &  3.832 &  3.796\\
$10^{7.0}\msun{}$ NSC &  2.449 & 0.59 & $2.179 \times 10^{-5}$ & $3.885 \times 10^{-6}$ & 0.1783 & $3.759 \times 10^{-5}$ & 0.08486 &  4.361 &  4.767\\
$10^{6.5}\msun{}$ NSC &  4.014 & 0.45 & $2.635 \times 10^{-5}$ & $2.705 \times 10^{-6}$ & 0.1027 & $4.87 \times 10^{-5}$ &  0.071 &   5.76 &  7.247\\
$10^{6.0}\msun{}$ NSC &  5.943 & 0.24 & $4.134 \times 10^{-5}$ & $2.49 \times 10^{-6}$ & 0.06023 & $9.202 \times 10^{-5}$ & 0.04152 &  7.929 &  11.05\\
$10^{5.5}\msun{}$ NSC &  7.678 & $\approx 0$ & $9.387 \times 10^{-5}$ & $3.466 \times 10^{-6}$ & 0.03692 & 0.0002763 & 0.02032 &  11.89 &  17.22\\
$10^{5.0}\msun{}$ NSC &  8.873 & $\approx 0$ & 0.000349 & $7.536 \times 10^{-6}$ & 0.02159 & 0.001501 & 0.00874 &  20.92 &  29.88\\
\hline
$10^{7.5}\msun{}$ bulge &  1.529 & 1.34 & $7.534 \times 10^{-6}$ & $4.389 \times 10^{-6}$ & 0.5825 & $1.026 \times 10^{-5}$ & 0.3421 &  1.364 &  1.324\\
$10^{7.0}\msun{}$ bulge &  2.449 & 2.09 & $2.853 \times 10^{-6}$ & $9.889 \times 10^{-7}$ & 0.3467 & $3.445 \times 10^{-6}$ & 0.2233 &  1.649 &  1.981\\
$10^{6.5}\msun{}$ bulge &  4.014 & 2.95 & $1.239 \times 10^{-6}$ & $2.602 \times 10^{-7}$ &   0.21 & $1.436 \times 10^{-6}$ & 0.1352 &   1.77 &  2.629\\
$10^{6.0}\msun{}$ bulge &  5.943 & 4.76 & $5.157 \times 10^{-7}$ & $7.45 \times 10^{-8}$ & 0.1445 & $5.785 \times 10^{-7}$ &  0.076 &  1.711 &  3.039\\
$10^{5.5}\msun{}$ bulge &  7.678 & 7.10 & $1.921 \times 10^{-7}$ & $2.204 \times 10^{-8}$ & 0.1147 & $2.099 \times 10^{-7}$ & 0.04024 &  1.578 &  3.195\\
$10^{5.0}\msun{}$ bulge &  8.873 & 10.98 & $6.384 \times 10^{-8}$ & $6.45 \times 10^{-9}$ &  0.101 & $6.843 \times 10^{-8}$ & 0.02043 &  1.458 &  3.226\\
		\hline
	\end{tabular}
\end{table*}

\begin{table*}
	\centering
	\caption{Same as Tab.~\ref{tab:rates0.03} assuming the maximum value of $\Delta m/m$ at which PTDEs can occur is 10 per cent. }
	\label{tab:rates0.1}
	\begin{tabular}{rccccccccc}
		\hline
		Model  & $\kappa$ & $q$ & P+TTDE  & TTDE & TTDE/P+TTDE & P+TTDE rep. & horizon/TTDE &  $\eta_{\rm TTDE}$ overest. &  $\eta=1$ overest. \\
		\hline
            $10^{7.5}\msun{}$ NSC &  1.237 & 0.26 & $1.999 \times 10^{-5}$ & $1.059 \times 10^{-5}$ & 0.5298 & $4.337 \times 10^{-5}$ & 0.07801 &   1.82 &  1.803\\
$10^{7.0}\msun{}$ NSC &  1.565 & 0.92 & $1.944 \times 10^{-5}$ & $9.255 \times 10^{-6}$ & 0.4761 & $2.232 \times 10^{-5}$ & 0.1224 &  1.831 &  2.001\\
$10^{6.5}\msun{}$ NSC &  2.003 & 0.90 & $2.089 \times 10^{-5}$ & $7.516 \times 10^{-6}$ & 0.3599 & $2.335 \times 10^{-5}$ & 0.08998 &  2.073 &  2.608\\
$10^{6.0}\msun{}$ NSC &  2.438 & 0.59 & $2.975 \times 10^{-5}$ & $8.028 \times 10^{-6}$ & 0.2698 & $3.446 \times 10^{-5}$ & 0.04938 &  2.459 &  3.426\\
$10^{5.5}\msun{}$ NSC &  2.771 & 0.16 & $6.48 \times 10^{-5}$ & $1.313 \times 10^{-5}$ & 0.2027 & $8.02 \times 10^{-5}$ & 0.02314 &  3.138 &  4.544\\
$10^{5.0}\msun{}$ NSC &  2.979 & $\approx 0$ & 0.0002427 & $3.499 \times 10^{-5}$ & 0.1442 & 0.0003325 & 0.00952 &  4.505 &  6.436\\
\hline
$10^{7.5}\msun{}$ bulge &  1.237 & 1.65 & $6.732 \times 10^{-6}$ & $5.136 \times 10^{-6}$ &  0.763 & $7.732 \times 10^{-6}$ & 0.3421 &  1.165 &  1.131\\
$10^{7.0}\msun{}$ bulge &  1.565 & 2.81 & $2.178 \times 10^{-6}$ & $1.287 \times 10^{-6}$ & 0.5909 & $2.273 \times 10^{-6}$ & 0.2258 &  1.267 &  1.522\\
$10^{6.5}\msun{}$ bulge &  2.003 & 5.00 & $7.698 \times 10^{-7}$ & $3.548 \times 10^{-7}$ & 0.4609 & $7.906 \times 10^{-7}$ & 0.1378 &  1.299 &  1.929\\
$10^{6.0}\msun{}$ bulge &  2.438 & 7.99 & $2.621 \times 10^{-7}$ & $9.988 \times 10^{-8}$ &  0.381 & $2.677 \times 10^{-7}$ & 0.07771 &  1.276 &  2.267\\
$10^{5.5}\msun{}$ bulge &  2.771 & 10.51 & $8.333 \times 10^{-8}$ & $2.819 \times 10^{-8}$ & 0.3383 & $8.473 \times 10^{-8}$ & 0.04093 &  1.233 &  2.498\\
$10^{5.0}\msun{}$ bulge &  2.979 & 13.08 & $2.486 \times 10^{-8}$ & $7.826 \times 10^{-9}$ & 0.3148 & $2.519 \times 10^{-8}$ & 0.02071 &  1.202 &  2.659\\ 
		\hline
	\end{tabular}
\end{table*}
\bsp	
\label{lastpage}
\end{document}